%Paper: cond-mat/9312003
%From: Fong Liu <fong@sbitp.itp.ucsb.edu>
%Date: Tue, 30 Nov 1993 18:19:31 -0800

% mail tex of paper start from line  1724
%
\expandafter\ifx\csname phyzzx\endcsname\relax
 \message{It is better to use PHYZZX format than to
          \string\input\space PHYZZX}\else
 \wlog{PHYZZX macros are already loaded and are not
          \string\input\space again}%
 \endinput \fi
\catcode`\@=11 % This allows us to modify PLAIN macros.
\let\rel@x=\relax
\let\n@expand=\relax
\def\pr@tect{\let\n@expand=\noexpand}
\let\protect=\pr@tect
\let\gl@bal=\global
\newfam\cpfam
\newdimen\b@gheight             \b@gheight=12pt
\newcount\f@ntkey               \f@ntkey=0
\def\f@m{\afterassignment\samef@nt\f@ntkey=}
\def\samef@nt{\fam=\f@ntkey \the\textfont\f@ntkey\rel@x}
\def\setstr@t{\setbox\strutbox=\hbox{\vrule height 0.85\b@gheight
                                depth 0.35\b@gheight width\z@ }}

\font\fourteenrm  =cmr10 scaled\magstep2
\font\twelverm    =cmr10 scaled\magstep1
\font\ninerm      =cmr9
\font\sixrm       =cmr6

\font\fourteenbf  =cmbx10 scaled\magstep3
\font\twelvebf    =cmbx10 scaled\magstep1
\font\ninebf      =cmbx9
\font\sixbf       =cmbx6
\font\seventeeni  =cmmi10 scaled\magstep3    \skewchar\seventeeni='177
\font\fourteeni   =cmmi10 scaled\magstep2     \skewchar\fourteeni='177
\font\twelvei     =cmmi10 scaled\magstep1       \skewchar\twelvei='177
\font\ninei       =cmmi9                          \skewchar\ninei='177
\font\sixi        =cmmi6                           \skewchar\sixi='177
\font\seventeensy =cmsy10 scaled\magstep3    \skewchar\seventeensy='60
\font\fourteensy  =cmsy10 scaled\magstep2     \skewchar\fourteensy='60
\font\twelvesy    =cmsy10 scaled\magstep1       \skewchar\twelvesy='60
\font\ninesy      =cmsy9                          \skewchar\ninesy='60
\font\sixsy       =cmsy6                           \skewchar\sixsy='60

\font\fourteenex  =cmex10 scaled\magstep2
\font\twelveex    =cmex10 scaled\magstep1

\font\fourteensl  =cmsl10 scaled\magstep2
\font\twelvesl    =cmsl10 scaled\magstep1
\font\ninesl      =cmsl9

\font\fourteenit  =cmti10 scaled\magstep2
\font\twelveit    =cmti10 scaled\magstep1
\font\nineit      =cmti9
\font\fourteentt  =cmtt10 scaled\magstep2
\font\twelvett    =cmtt10 scaled\magstep1
\font\fourteencp  =cmcsc10 scaled\magstep2
\font\twelvecp    =cmcsc10 scaled\magstep1
\font\tencp       =cmcsc10
%
%%%%%%%%%%%%%%%%%%%%%%%%%%%%%%%%%%%%%%%%%%%%%%%%%%%%%%%%%%
%
\def\fourteenf@nts{\relax
    \textfont0=\fourteenrm          \scriptfont0=\tenrm
      \scriptscriptfont0=\sevenrm
    \textfont1=\fourteeni           \scriptfont1=\teni
      \scriptscriptfont1=\seveni
    \textfont2=\fourteensy          \scriptfont2=\tensy
      \scriptscriptfont2=\sevensy
    \textfont3=\fourteenex          \scriptfont3=\twelveex
      \scriptscriptfont3=\tenex
    \textfont\itfam=\fourteenit     \scriptfont\itfam=\tenit
    \textfont\slfam=\fourteensl     \scriptfont\slfam=\tensl
    \textfont\bffam=\fourteenbf     \scriptfont\bffam=\tenbf
      \scriptscriptfont\bffam=\sevenbf
    \textfont\ttfam=\fourteentt
    \textfont\cpfam=\fourteencp }
\def\twelvef@nts{\relax
    \textfont0=\twelverm          \scriptfont0=\ninerm
      \scriptscriptfont0=\sixrm
    \textfont1=\twelvei           \scriptfont1=\ninei
      \scriptscriptfont1=\sixi
    \textfont2=\twelvesy           \scriptfont2=\ninesy
      \scriptscriptfont2=\sixsy
    \textfont3=\twelveex          \scriptfont3=\tenex
      \scriptscriptfont3=\tenex
    \textfont\itfam=\twelveit     \scriptfont\itfam=\nineit
    \textfont\slfam=\twelvesl     \scriptfont\slfam=\ninesl
    \textfont\bffam=\twelvebf     \scriptfont\bffam=\ninebf
      \scriptscriptfont\bffam=\sixbf
    \textfont\ttfam=\twelvett
    \textfont\cpfam=\twelvecp }
\def\tenf@nts{\relax
    \textfont0=\tenrm          \scriptfont0=\sevenrm
      \scriptscriptfont0=\fiverm
    \textfont1=\teni           \scriptfont1=\seveni
      \scriptscriptfont1=\fivei
    \textfont2=\tensy          \scriptfont2=\sevensy
      \scriptscriptfont2=\fivesy
    \textfont3=\tenex          \scriptfont3=\tenex
      \scriptscriptfont3=\tenex
    \textfont\itfam=\tenit     \scriptfont\itfam=\seveni  % no \sevenit
    \textfont\slfam=\tensl     \scriptfont\slfam=\sevenrm % no \sevensl
    \textfont\bffam=\tenbf     \scriptfont\bffam=\sevenbf
      \scriptscriptfont\bffam=\fivebf
    \textfont\ttfam=\tentt
    \textfont\cpfam=\tencp }
%
%%%%%%%%%%%%%%%%%%%%%%%%%%%%%%%%%%%%%%%%%%%%%%%%%%%%%%%%%%
%

% Actual font definitions are kept in a separate file
% to facilitate font substitution.
%
\def\rm{\n@expand\f@m0 }
\def\mit{\n@expand\f@m1 }         
\def\cal{\n@expand\f@m2 }
\def\it{\n@expand\f@m\itfam}
\def\sl{\n@expand\f@m\slfam}
\def\bf{\n@expand\f@m\bffam}
\def\tt{\n@expand\f@m\ttfam}
\def\caps{\n@expand\f@m\cpfam}    
\def\em@{\rel@x\ifnum\f@ntkey=0 \it \else
        \ifnum\f@ntkey=\bffam \it \else \rm \fi \fi }
\def\em{\n@expand\em@}
\def\fourteenpoint{\fourteenf@nts \samef@nt \b@gheight=14pt \setstr@t }
\def\twelvepoint{\twelvef@nts \samef@nt \b@gheight=12pt \setstr@t }
\def\tenpoint{\tenf@nts \samef@nt \b@gheight=10pt \setstr@t }
\normalbaselineskip = 20pt plus 0.2pt minus 0.1pt
\normallineskip = 1.5pt plus 0.1pt minus 0.1pt
\normallineskiplimit = 1.5pt
\newskip\normaldisplayskip
\normaldisplayskip = 20pt plus 5pt minus 10pt
\newskip\normaldispshortskip
\normaldispshortskip = 6pt plus 5pt
\newskip\normalparskip
\normalparskip = 6pt plus 2pt minus 1pt
\newskip\skipregister
\skipregister = 5pt plus 2pt minus 1.5pt
\newif\ifsingl@
\newif\ifdoubl@
\newif\iftwelv@  \twelv@true
\def\singlespace{\singl@true\doubl@false\spaces@t}
\def\doublespace{\singl@false\doubl@true\spaces@t}
\def\normalspace{\singl@false\doubl@false\spaces@t}
\def\Tenpoint{\tenpoint\twelv@false\spaces@t}
\def\Twelvepoint{\twelvepoint\twelv@true\spaces@t}
\def\spaces@t{\rel@x
      \iftwelv@ \ifsingl@\subspaces@t3:4;\else\subspaces@t1:1;\fi
       \else \ifsingl@\subspaces@t3:5;\else\subspaces@t4:5;\fi \fi
      \ifdoubl@ \multiply\baselineskip by 5
         \divide\baselineskip by 4 \fi }
\def\subspaces@t#1:#2;{
      \baselineskip = \normalbaselineskip
      \multiply\baselineskip by #1 \divide\baselineskip by #2
      \lineskip = \normallineskip
      \multiply\lineskip by #1 \divide\lineskip by #2
      \lineskiplimit = \normallineskiplimit
      \multiply\lineskiplimit by #1 \divide\lineskiplimit by #2
      \parskip = \normalparskip
      \multiply\parskip by #1 \divide\parskip by #2
      \abovedisplayskip = \normaldisplayskip
      \multiply\abovedisplayskip by #1 \divide\abovedisplayskip by #2
      \belowdisplayskip = \abovedisplayskip
      \abovedisplayshortskip = \normaldispshortskip
      \multiply\abovedisplayshortskip by #1
        \divide\abovedisplayshortskip by #2
      \belowdisplayshortskip = \abovedisplayshortskip
      \advance\belowdisplayshortskip by \belowdisplayskip
      \divide\belowdisplayshortskip by 2
      \smallskipamount = \skipregister
      \multiply\smallskipamount by #1 \divide\smallskipamount by #2
      \medskipamount = \smallskipamount \multiply\medskipamount by 2
      \bigskipamount = \smallskipamount \multiply\bigskipamount by 4 }
\def\normalbaselines{ \baselineskip=\normalbaselineskip
   \lineskip=\normallineskip \lineskiplimit=\normallineskip
   \iftwelv@\else \multiply\baselineskip by 4 \divide\baselineskip by 5
     \multiply\lineskiplimit by 4 \divide\lineskiplimit by 5
     \multiply\lineskip by 4 \divide\lineskip by 5 \fi }
\Twelvepoint  % That's the default
\interlinepenalty=50
\interfootnotelinepenalty=5000
\predisplaypenalty=9000
\postdisplaypenalty=500
\hfuzz=1pt
\vfuzz=0.2pt
\newdimen\HOFFSET  \HOFFSET=0pt
\newdimen\VOFFSET  \VOFFSET=0pt
\newdimen\HSWING   \HSWING=0pt
\dimen\footins=8in
%
%%%%%%%%%%%%%%%%%%%%%%%%%%%%%%%%%%%%%%%%%%%%%%%%%%%%%%%%%%%%%%%%%%%%%%%%
%
%   Next, I define output routines, footnotes & related stuff.
%
\newskip\pagebottomfiller
\pagebottomfiller=\z@ plus \z@ minus \z@
\def\pagecontents{
   \ifvoid\topins\else\unvbox\topins\vskip\skip\topins\fi
   \dimen@ = \dp255 \unvbox255
   \vskip\pagebottomfiller
   \ifvoid\footins\else\vskip\skip\footins\footrule\unvbox\footins\fi
   \ifr@ggedbottom \kern-\dimen@ \vfil \fi }
\def\makeheadline{\vbox to 0pt{ \skip@=\topskip
      \advance\skip@ by -12pt \advance\skip@ by -2\normalbaselineskip
      \vskip\skip@ \line{\vbox to 12pt{}\the\headline} \vss
      }\nointerlineskip}
\def\makefootline{\baselineskip = 1.5\normalbaselineskip
                 \line{\the\footline}}
\newif\iffrontpage
\newif\ifp@genum
\def\nopagenumbers{\p@genumfalse}
\def\pagenumbers{\p@genumtrue}
\pagenumbers
\newtoks\paperheadline
\newtoks\paperfootline
\newtoks\letterheadline
\newtoks\letterfootline
\newtoks\letterinfo
\newtoks\date
\paperheadline={\hfil}
\paperfootline={\hss\iffrontpage\else\ifp@genum\tenrm\folio\hss\fi\fi}
\letterheadline{\iffrontpage \hfil \else
    \rm \ifp@genum page~~\folio\fi \hfil\the\date \fi}
\letterfootline={\iffrontpage\the\letterinfo\else\hfil\fi}
\letterinfo={\hfil}
\def\monthname{\rel@x\ifcase\month 0/\or January\or February\or
   March\or April\or May\or June\or July\or August\or September\or
   October\or November\or December\else\number\month/\fi}
\def\today{\monthname~\number\day, \number\year}
\date={\today}
\headline=\paperheadline % The default is
\footline=\paperfootline % \papers
\countdef\pageno=1      \countdef\pagen@=0
\countdef\pagenumber=1  \pagenumber=1
\def\advancepageno{\gl@bal\advance\pagen@ by 1
   \ifnum\pagenumber<0 \gl@bal\advance\pagenumber by -1
    \else\gl@bal\advance\pagenumber by 1 \fi
    \gl@bal\frontpagefalse  \swing@ }
\def\folio{\ifnum\pagenumber<0 \romannumeral-\pagenumber
           \else \number\pagenumber \fi }
\def\swing@{\ifodd\pagenumber \gl@bal\advance\hoffset by -\HSWING
             \else \gl@bal\advance\hoffset by \HSWING \fi }
\def\footrule{\dimen@=\prevdepth\nointerlineskip
   \vbox to 0pt{\vskip -0.25\baselineskip \hrule width 0.35\hsize \vss}
   \prevdepth=\dimen@ }
\let\footnotespecial=\rel@x
\newdimen\footindent
\footindent=24pt
\def\Textindent#1{\noindent\llap{#1\enspace}\ignorespaces}
\def\Vfootnote#1{\insert\footins\bgroup
   \interlinepenalty=\interfootnotelinepenalty \floatingpenalty=20000
   \singl@true\doubl@false\Tenpoint
   \splittopskip=\ht\strutbox \boxmaxdepth=\dp\strutbox
   \leftskip=\footindent \rightskip=\z@skip
   \parindent=0.5\footindent \parfillskip=0pt plus 1fil
   \spaceskip=\z@skip \xspaceskip=\z@skip \footnotespecial
   \Textindent{#1}\footstrut\futurelet\next\fo@t}

\def\vfootnote#1{\Vfootnote{${#1}$}}
\def\footnote#1{\attach{#1}\vfootnote{#1}}

\let\footsymbol=\star
\newcount\lastf@@t           \lastf@@t=-1
\newcount\footsymbolcount    \footsymbolcount=0
\newif\ifPhysRev
\def\bumpfootsymbolcount{\rel@x
   \iffrontpage \bumpfootsymbolpos \else \advance\lastf@@t by 1
     \ifPhysRev \bumpfootsymbolneg \else \bumpfootsymbolpos \fi \fi
   \gl@bal\lastf@@t=\pagen@ }
\def\bumpfootsymbolpos{\ifnum\footsymbolcount <0
                            \gl@bal\footsymbolcount =0 \fi
    \ifnum\lastf@@t<\pagen@ \gl@bal\footsymbolcount=0
     \else \gl@bal\advance\footsymbolcount by 1 \fi }
\def\bumpfootsymbolneg{\ifnum\footsymbolcount >0
             \gl@bal\footsymbolcount =0 \fi
         \gl@bal\advance\footsymbolcount by -1 }
\def\fd@f#1 {\xdef\footsymbol{\mathchar"#1 }}
\def\generatefootsymbol{\ifcase\footsymbolcount \fd@f 13F \or \fd@f 279
        \or \fd@f 27A \or \fd@f 278 \or \fd@f 27B \else
        \ifnum\footsymbolcount <0 \fd@f{023 \number-\footsymbolcount }
         \else \fd@f 203 {\loop \ifnum\footsymbolcount >5
                \fd@f{203 \footsymbol } \advance\footsymbolcount by -1
                \repeat }\fi \fi }

\def\nonfrenchspacing{\sfcode`\.=3001 \sfcode`\!=3000 \sfcode`\?=3000
        \sfcode`\:=2000 \sfcode`\;=1500 \sfcode`\,=1251 }
\nonfrenchspacing
\newdimen\d@twidth
{\setbox0=\hbox{s.} \gl@bal\d@twidth=\wd0 \setbox0=\hbox{s}
        \gl@bal\advance\d@twidth by -\wd0 }
\def\removehglue{\loop \unskip \ifdim\lastskip >\z@ \repeat }
\def\roll@ver#1{\removehglue \nobreak \count255 =\spacefactor \dimen@=\z@
        \ifnum\count255 =3001 \dimen@=\d@twidth \fi
        \ifnum\count255 =1251 \dimen@=\d@twidth \fi
    \iftwelv@ \kern-\dimen@ \else \kern-0.83\dimen@ \fi
   #1\spacefactor=\count255 }
\def\step@ver#1{\rel@x \ifmmode #1\else \ifhmode
        \roll@ver{${}#1$}\else {\setbox0=\hbox{${}#1$}}\fi\fi }
\def\attach#1{\step@ver{\strut^{\mkern 2mu #1} }}
%
%%%%%%%%%%%%%%%%%%%%%%%%%%%%%%%%%%%%%%%%%%%%%%%%%%%%%%%%%%%%%%%%%%%%%%%%
%
%   Here come chapter, section, subsection & appendix macros.
%
\newcount\chapternumber      \chapternumber=0
\newcount\sectionnumber      \sectionnumber=0
\newcount\equanumber         \equanumber=0
\let\chapterlabel=\rel@x
\let\sectionlabel=\rel@x
\newtoks\chapterstyle        \chapterstyle={\Number}
\newtoks\sectionstyle        \sectionstyle={\Number}
\newskip\chapterskip         \chapterskip=\bigskipamount
\newskip\sectionskip         \sectionskip=\medskipamount
\newskip\headskip            \headskip=8pt plus 3pt minus 3pt
\newdimen\chapterminspace    \chapterminspace=15pc
\newdimen\sectionminspace    \sectionminspace=10pc
\newdimen\referenceminspace  \referenceminspace=20pc
\newif\ifcn@                 \cn@true
\newif\ifcn@@                \cn@@false
\def\numberedchapters{\cn@true}
\def\unnumberedchapters{\cn@false\sequentialequations}
\def\chapterreset{\gl@bal\advance\chapternumber by 1
   \ifnum\equanumber<0 \else\gl@bal\equanumber=0\fi
   \sectionnumber=0 \let\sectionlabel=\rel@x
   \ifcn@ \gl@bal\cn@@true {\pr@tect
       \xdef\chapterlabel{\the\chapterstyle{\the\chapternumber}}}%
    \else \gl@bal\cn@@false \gdef\chapterlabel{\rel@x}\fi }
\def\@alpha#1{\count255='140 \advance\count255 by #1\char\count255}
 \def\alphabetic{\n@expand\@alpha}
\def\@Alpha#1{\count255='100 \advance\count255 by #1\char\count255}
 \def\Alphabetic{\n@expand\@Alpha}
\def\@Roman#1{\uppercase\expandafter{\romannumeral #1}}
 \def\Roman{\n@expand\@Roman}
\def\@roman#1{\romannumeral #1}    \def\roman{\n@expand\@roman}
\def\@number#1{\number #1}         \def\Number{\n@expand\@number}
\def\BLANK#1{\rel@x}               
\def\titleparagraphs{\interlinepenalty=9999
     \leftskip=0.03\hsize plus 0.22\hsize minus 0.03\hsize
     \rightskip=\leftskip \parfillskip=0pt
     \hyphenpenalty=9000 \exhyphenpenalty=9000
     \tolerance=9999 \pretolerance=9000
     \spaceskip=0.333em \xspaceskip=0.5em }
\def\titlestyle#1{\par\begingroup \titleparagraphs
     \iftwelv@\fourteenpoint\else\twelvepoint\fi
   \noindent #1\par\endgroup }
\def\spacecheck#1{\dimen@=\pagegoal\advance\dimen@ by -\pagetotal
   \ifdim\dimen@<#1 \ifdim\dimen@>0pt \vfil\break \fi\fi}
\def\chapter#1{\par \penalty-300 \vskip\chapterskip
   \spacecheck\chapterminspace
   \chapterreset \titlestyle{\ifcn@@\chapterlabel.~\fi #1}
   \nobreak\vskip\headskip \penalty 30000
   {\pr@tect\wlog{\string\chapter\space \chapterlabel}} }

%
%\def\section#1{\par \ifnum\lastpenalty=30000\else
%   \penalty-200\vskip\sectionskip \spacecheck\sectionminspace\fi
%   \gl@bal\advance\sectionnumber by 1
%   {\pr@tect
%  \ifcn@@\expandafter\toks@\expandafter{\chapterlabel.}\else\toks@={}\fi
%   \xdef\sectionlabel{\the\toks@\the\sectionstyle{\the\sectionnumber}}%
%   \wlog{\string\section\space \sectionlabel}}%
%   \noindent {\caps\enspace\sectionlabel.~~#1}\par
%   \nobreak\vskip\headskip \penalty 30000 }

%  I am testing the section fonts, i changed \caps

\def\section#1{\par \ifnum\lastpenalty=30000\else
   \penalty-200\vskip\sectionskip \spacecheck\sectionminspace\fi
   \gl@bal\advance\sectionnumber by 1
   {\pr@tect
   \ifcn@@\expandafter\toks@\expandafter{\chapterlabel.}\else\toks@={}\fi
   \xdef\sectionlabel{\the\toks@\the\sectionstyle{\the\sectionnumber}}%
   \wlog{\string\section\space \sectionlabel}}%
   \noindent {\bf\enspace\sectionlabel.~~#1}\par
   \nobreak\vskip\headskip \penalty 30000 }

\def\subsection#1{\par
   \ifnum\the\lastpenalty=30000\else \penalty-100\smallskip \fi
   \noindent\undertext{#1}\enspace \vadjust{\penalty5000}}

\def\undertext#1{\vtop{\hbox{#1}\kern 1pt \hrule}}

\def\ack{\subsection{Acknowledgements:}}
\def\APPENDIX#1#2{\par\penalty-300\vskip\chapterskip
   \spacecheck\chapterminspace \chapterreset \xdef\chapterlabel{#1}
   \titlestyle{APPENDIX #2} \nobreak\vskip\headskip \penalty 30000
   \wlog{\string\Appendix~\chapterlabel} }
\def\Appendix#1{\APPENDIX{#1}{#1}}
\def\appendix{\APPENDIX{A}{}}
%
%%%%%%%%%%%%%%%%%%%%%%%%%%%%%%%%%%%%%%%%%%%%%%%%%%%%%%%%%%%%%%%%%%%%%%%%
%
%   Here come macros for equation numbering.
%
%\def\eqname#1{\rel@x {\pr@tect
%  \ifnum\equanumber<0 \xdef#1{{\rm(\number-\equanumber)}}%
%     \gl@bal\advance\equanumber by -1
%  \else \gl@bal\advance\equanumber by 1
%   \ifcn@@ \toks@=\expandafter{\chapterlabel.}\else\toks@={}\fi
%   \xdef#1{{\rm(\the\toks@\number\equanumber)}}\fi #1}}
%
% Modified for Vanilla
%
\def\eqname#1{\rel@x {\pr@tect
  \ifnum\equanumber<0 \xdef#1{{(\number-\equanumber)}}%
     \gl@bal\advance\equanumber by -1
  \else \gl@bal\advance\equanumber by 1
   \ifcn@@ \toks@=\expandafter{\chapterlabel.}\else\toks@={}\fi
   \xdef#1{{(\the\toks@\number\equanumber)}}\fi #1}}

\def\eq{\eqname\?}
\def\eqn{\eqno\eqname}

\def\eqinsert#1{\noalign{\dimen@=\prevdepth \nointerlineskip
   \setbox0=\hbox to\displaywidth{\hfil #1}
   \vbox to 0pt{\kern 0.5\baselineskip\hbox{$\!\box0\!$}\vss}
   \prevdepth=\dimen@}}
%

%
%%%%%%%%%%%%%%%%%%%%%%%%%%%%%%%%%%%%%%%%%%%%%%%%%%%%%%%%%%%%%%%%%%%%%%%%
%   Here come items and lists
%
\def\GENITEM#1;#2{\par \hangafter=0 \hangindent=#1
    \Textindent{$ #2 $}\ignorespaces}
\outer\def\newitem#1=#2;{\gdef#1{\GENITEM #2;}}

\newdimen\itemsize                \itemsize=30pt
\newitem\item=1\itemsize;
\newitem\sitem=1.75\itemsize;     
\newitem\ssitem=2.5\itemsize;     
\outer\def\newlist#1=#2&#3&#4;{\toks0={#2}\toks1={#3}%
   \count255=\escapechar \escapechar=-1
   \alloc@0\list\countdef\insc@unt\listcount     \listcount=0
   \edef#1{\par
      \countdef\listcount=\the\allocationnumber
      \advance\listcount by 1
      \hangafter=0 \hangindent=#4
      \Textindent{\the\toks0{\listcount}\the\toks1}}
   \expandafter\expandafter\expandafter
    \edef\c@t#1{begin}{\par
      \countdef\listcount=\the\allocationnumber \listcount=1
      \hangafter=0 \hangindent=#4
      \Textindent{\the\toks0{\listcount}\the\toks1}}
   \expandafter\expandafter\expandafter
    \edef\c@t#1{con}{\par \hangafter=0 \hangindent=#4 \noindent}
   \escapechar=\count255}
\def\c@t#1#2{\csname\string#1#2\endcsname}
\newlist\point=\Number&.&1.0\itemsize;
\newlist\subpoint=(\alphabetic&)&1.75\itemsize;
\newlist\subsubpoint=(\roman&)&2.5\itemsize;
%

%
%%%%%%%%%%%%%%%%%%%%%%%%%%%%%%%%%%%%%%%%%%%%%%%%%%%%%%%%%%%%%%%%%%%%%%%%
%
%   Here come macros for references, figures & tables.
%
% % % % % % % % % % % % % % % % % % % % % % % % % % % % % % % % % % % %
%%  First, references.
%
\newcount\referencecount     \referencecount=0
\newcount\lastrefsbegincount \lastrefsbegincount=0
\newif\ifreferenceopen       \newwrite\referencewrite
\newdimen\refindent          \refindent=30pt
\def\normalrefmark#1{\attach{\scriptscriptstyle [ #1 ] }}
\let\PRrefmark=\attach
\def\NPrefmark#1{\step@ver{{\;[#1]}}}
\def\refmark#1{\rel@x\ifPhysRev\PRrefmark{#1}\else\normalrefmark{#1}\fi}
\def\refend@{\refmark{\number\referencecount}}
\def\refend{\refend@{}\space }
\def\refsend{\refmark{\count255=\referencecount
   \advance\count255 by-\lastrefsbegincount
   \ifcase\count255 \number\referencecount
   \or \number\lastrefsbegincount,\number\referencecount
   \else \number\lastrefsbegincount-\number\referencecount \fi}\space }
\def\REFNUM#1{\rel@x \gl@bal\advance\referencecount by 1
    \xdef#1{\the\referencecount }}
\def\Refnum#1{\REFNUM #1\refend@ } \let\refnum=\Refnum
\def\REF#1{\REFNUM #1\R@FWRITE\ignorespaces}
\def\Ref#1{\Refnum #1\REFWRITE }
\def\ref{\Ref\?}
\def\REFS#1{\REFNUM #1\gl@bal\lastrefsbegincount=\referencecount
    \REFWRITE }

       \let\REFSCON=\REF
\def\r@fitem#1{\par \hangafter=0 \hangindent=\refindent \Textindent{#1}}
\def\refitem#1{\r@fitem{#1.}}
\def\NPrefitem#1{\r@fitem{[#1]}}
\def\NPrefs{\let\refmark=\NPrefmark \let\refitem=NPrefitem}
\def\REFWRITE{\R@FWRITE\rel@x }
\def\R@FWRITE#1{\ifreferenceopen \else \gl@bal\referenceopentrue
     \immediate\openout\referencewrite=\jobname.refs
     \toks@={\begingroup \refoutspecials \catcode`\^^M=10 }%
     \immediate\write\referencewrite{\the\toks@}\fi
    \immediate\write\referencewrite{\noexpand\refitem %
                                    {\the\referencecount}}%
    \p@rse@ndwrite \referencewrite #1}
\begingroup
 \catcode`\^^M=\active \let^^M=\relax %
 \gdef\p@rse@ndwrite#1#2{\begingroup \catcode`\^^M=12 \newlinechar=`\^^M%
         \chardef\rw@write=#1\sc@nlines#2}%
 \gdef\sc@nlines#1#2{\sc@n@line \g@rbage #2^^M\endsc@n \endgroup #1}%
 \gdef\sc@n@line#1^^M{\expandafter\toks@\expandafter{\deg@rbage #1}%
         \immediate\write\rw@write{\the\toks@}%
         \futurelet\n@xt \sc@ntest }%
\endgroup
\def\sc@ntest{\ifx\n@xt\endsc@n \let\n@xt=\rel@x
       \else \let\n@xt=\sc@n@notherline \fi \n@xt }
\def\sc@n@notherline{\sc@n@line \g@rbage }
\def\deg@rbage#1{}
\let\g@rbage=\relax    \let\endsc@n=\relax
\def\refout{\par\penalty-400\vskip\chapterskip
   \spacecheck\referenceminspace
   \ifreferenceopen \Closeout\referencewrite \referenceopenfalse \fi
   \line{\fourteenrm\hfil REFERENCES\hfil}\vskip\headskip
   \input \jobname.refs
   }

\def\myrefout{\par\penalty-400
   \spacecheck\referenceminspace
    \ifreferenceopen \Closeout\referencewrite \referenceopenfalse \fi
   \input \jobname.refs
   }

\def\refoutspecials{\sfcode`\.=1000 \interlinepenalty=1000
         \rightskip=\z@ plus 1em minus \z@ }
\def\Closeout#1{\toks0={\par\endgroup}\immediate\write#1{\the\toks0}%
   \immediate\closeout#1}
%
% % % % % % % % % % % % % % % % % % % % % % % % % % % % % % % % % % % %
%%  Next, figure captions and table captions.
%
\newcount\figurecount     \figurecount=0
\newcount\tablecount      \tablecount=0
\newif\iffigureopen       \newwrite\figurewrite
\newif\iftableopen        \newwrite\tablewrite
\def\FIGNUM#1{\rel@x \gl@bal\advance\figurecount by 1
    \xdef#1{\the\figurecount}}
\def\FIGURE#1{\FIGNUM #1\F@GWRITE\ignorespaces }
\let\FIG=\FIGURE

\def\figitem#1{\r@fitem{#1)}}
\def\FIGWRITE{\F@GWRITE\rel@x }
\def\TABNUM#1{\rel@x \gl@bal\advance\tablecount by 1
    \xdef#1{\the\tablecount}}
\def\TABLE#1{\TABNUM #1\T@BWRITE\ignorespaces }

\def\tabitem#1{\r@fitem{#1:}}
\def\TABWRITE{\T@BWRITE\rel@x }
\def\F@GWRITE#1{\iffigureopen \else \gl@bal\figureopentrue
     \immediate\openout\figurewrite=\jobname.figs
     \toks@={\begingroup \catcode`\^^M=10 }%
     \immediate\write\figurewrite{\the\toks@}\fi
    \immediate\write\figurewrite{\noexpand\figitem %
                                 {\the\figurecount}}%
    \p@rse@ndwrite \figurewrite #1}
\def\T@BWRITE#1{\iftableopen \else \gl@bal\tableopentrue
     \immediate\openout\tablewrite=\jobname.tabs
     \toks@={\begingroup \catcode`\^^M=10 }%
     \immediate\write\tablewrite{\the\toks@}\fi
    \immediate\write\tablewrite{\noexpand\tabitem %
                                 {\the\tablecount}}%
    \p@rse@ndwrite \tablewrite #1}
\def\figout{\par\penalty-400
   \vskip\chapterskip\spacecheck\referenceminspace
   \iffigureopen \Closeout\figurewrite \figureopenfalse \fi
   \line{\fourteenrm\hfil FIGURE CAPTIONS\hfil}\vskip\headskip
   \input \jobname.figs
   }
\def\tabout{\par\penalty-400
   \vskip\chapterskip\spacecheck\referenceminspace
   \iftableopen \Closeout\tablewrite \tableopenfalse \fi
   \line{\fourteenrm\hfil TABLE CAPTIONS\hfil}\vskip\headskip
   \input \jobname.tabs
   }
%
% % % % % % % % % % % % % % % % % % % % % % % % % % % % % % % % % % % %
%%  Finally, inserted figures.
%
%
\newbox\picturebox
\def\p@cht{\ht\picturebox }
\def\p@cwd{\wd\picturebox }
\def\p@cdp{\dp\picturebox }
\newdimen\xshift
\newdimen\yshift
\newdimen\captionwidth
\newskip\captionskip
\captionskip=15pt plus 5pt minus 3pt
\def\fullwidth{\captionwidth=\hsize }
\newtoks\Caption
\newif\ifcaptioned
\newif\ifselfcaptioned
\def\caption{\captionedtrue \Caption }
\newcount\linesabove
\newif\iffileexists
\newtoks\picfilename
\def\fil@#1 {\fileexiststrue \picfilename={#1}}
\def\file#1{\if=#1\let\n@xt=\fil@ \else \def\n@xt{\fil@ #1}\fi \n@xt }
\def\pl@t{\begingroup \pr@tect
    \setbox\picturebox=\hbox{}\fileexistsfalse
    \let\height=\p@cht \let\width=\p@cwd \let\depth=\p@cdp
    \xshift=\z@ \yshift=\z@ \captionwidth=\z@
    \Caption={}\captionedfalse
    \linesabove =0 \picturedefault }
\def\plot{\pl@t \selfcaptionedfalse }
\def\Picture#1{\gl@bal\advance\figurecount by 1
    \xdef#1{\the\figurecount}\pl@t \selfcaptionedtrue }

\def\s@vepicture{\iffileexists \parsefilename \redopicturebox \fi
   \ifdim\captionwidth>\z@ \else \captionwidth=\p@cwd \fi
   \xdef\lastpicture{\iffileexists
        \setbox0=\hbox{\raise\the\yshift \vbox{%
              \moveright\the\xshift\hbox{\picturedefinition}}}%
        \else \setbox0=\hbox{}\fi
         \ht0=\the\p@cht \wd0=\the\p@cwd \dp0=\the\p@cdp
         \vbox{\hsize=\the\captionwidth \line{\hss\box0 \hss }%
              \ifcaptioned \vskip\the\captionskip \noexpand\Tenpoint
                \ifselfcaptioned Figure~\the\figurecount.\enspace \fi
                \the\Caption \fi }}%
    \endgroup }
\let\endpicture=\s@vepicture
\def\savepicture#1{\s@vepicture \global\let#1=\lastpicture }
\def\displaypicture{\fullwidth \s@vepicture $$\lastpicture $${}}
\def\toppicture{\fullwidth \s@vepicture \topinsert
    \lastpicture \medskip \endinsert }
\def\midpicture{\fullwidth \s@vepicture \midinsert
    \lastpicture \endinsert }
%
%  Wraparound macros - a try.
%
\def\leftpicture{\pres@tpicture
    \dimen@i=\hsize \advance\dimen@i by -\dimen@ii
    \setbox\picturebox=\hbox to \hsize {\box0 \hss }%
    \wr@paround }
\def\rightpicture{\pres@tpicture
    \dimen@i=\z@
    \setbox\picturebox=\hbox to \hsize {\hss \box0 }%
    \wr@paround }
\def\pres@tpicture{\gl@bal\linesabove=\linesabove
    \s@vepicture \setbox\picturebox=\vbox{
         \kern \linesabove\baselineskip \kern 0.3\baselineskip
         \lastpicture \kern 0.3\baselineskip }%
    \dimen@=\p@cht \dimen@i=\dimen@
    \advance\dimen@i by \pagetotal
    \par \ifdim\dimen@i>\pagegoal \vfil\break \fi
    \dimen@ii=\hsize
    \advance\dimen@ii by -\parindent \advance\dimen@ii by -\p@cwd
    \setbox0=\vbox to\z@{\kern-\baselineskip \unvbox\picturebox \vss }}
\def\wr@paround{\Caption={}\count255=1
    \loop \ifnum \linesabove >0
         \advance\linesabove by -1 \advance\count255 by 1
         \advance\dimen@ by -\baselineskip
         \expandafter\Caption \expandafter{\the\Caption \z@ \hsize }%
      \repeat
    \loop \ifdim \dimen@ >\z@
         \advance\count255 by 1 \advance\dimen@ by -\baselineskip
         \expandafter\Caption \expandafter{%
             \the\Caption \dimen@i \dimen@ii }%
      \repeat
    \edef\n@xt{\parshape=\the\count255 \the\Caption \z@ \hsize }%
    \par\noindent \n@xt \strut \vadjust{\box\picturebox }}
\let\picturedefault=\relax
\let\parsefilename=\relax
\def\redopicturebox{\let\picturedefinition=\rel@x
   \errhelp=\disabledpictures
   \errmessage{This version of TeX cannot handle pictures.  Sorry.}}
\newhelp\disabledpictures
     {You will get a blank box in place of your picture.}
%
%
%
% Above definitions of \parsefilename and \redopicturebox
% are dumb defaults.  Actual definition are system dependent;
% you will probably find them in your PHYZZX.LOCAL file.
%
% The example below is used at Princeton.
%
%\def\parsefilename{\expandafter\p@rse\the\picfilename.\endp@rse }
%\def\p@rse#1.#2\endp@rse{\if"#2"\expandafter\picfilename
%        \expandafter{\the\picfilename.fig}\fi }
%
%\newread\pictureread
%\def\redopicturebox{\expandafter\openin\expandafter\pictureread
%                    \the\picfilename
%   \ifeof\pictureread \errhelp=\disabledpictures
%     \edef\n@xt{\errmessage={Cannot find file \the\picfilename}\n@xt
%     \let\pictureboxdefinition=\relax \else
%    \read\pictureread to\y@p \read\pictureread to\y@p
%    \read\pictureread to\x@p \read\pictureread to\y@m
%    \read\pictureread to\x@m \closein\pictureread
%    \p@cht=\y@p truein \advance\p@cht by -\y@m truein
%    \advance\yshift by \y@p truein
%    \p@cwd=\x@p truein \advance\p@cwd by -\x@m truein
%    \advance\xshift by \x@m truein
%    \edef\picturedefinition{\special{pos,inc=\the\picfilename}}%
%    \fi }
%
%
%%%%%%%%%%%%%%%%%%%%%%%%%%%%%%%%%%%%%%%%%%%%%%%%%%%%%%%%%%%%%%%%%%%%%%%%
%
%   Here come macros for memos & letters.
%
\def\FRONTPAGE{\ifvoid255\else\vfill\penalty-20000\fi
   \gl@bal\pagenumber=1     \gl@bal\chapternumber=0
   \gl@bal\equanumber=0     \gl@bal\sectionnumber=0
   \gl@bal\referencecount=0 \gl@bal\figurecount=0
   \gl@bal\tablecount=0     \gl@bal\frontpagetrue
   \gl@bal\lastf@@t=0       \gl@bal\footsymbolcount=0
   \gl@bal\cn@@false }

\def\papers{\papersize\headline=\paperheadline\footline=\paperfootline}
\def\papersize{\hsize=35pc \vsize=50pc \hoffset=0pc \voffset=1pc
   \advance\hoffset by\HOFFSET \advance\voffset by\VOFFSET
   \pagebottomfiller=0pc
   \skip\footins=\bigskipamount \normalspace }
\papers  %  This is the default
%
% % % % % % % % % % % % % % % % % % % % % % % % % % % % % % % % % % % %
%
\newskip\lettertopskip       \lettertopskip=20pt plus 50pt
\newskip\letterbottomskip    \letterbottomskip=\z@ plus 100pt
\newskip\signatureskip       \signatureskip=40pt plus 3pt
\def\lettersize{\hsize=6.5in \vsize=8.5in \hoffset=0in \voffset=0.5in
   \advance\hoffset by\HOFFSET \advance\voffset by\VOFFSET
   \pagebottomfiller=\letterbottomskip
   \skip\footins=\smallskipamount \multiply\skip\footins by 3
   \singlespace }
\def\MEMO{\lettersize \headline=\letterheadline \footline={\hfil }%
   \let\rule=\memorule \FRONTPAGE \memohead }

\def\memodate{\afterassignment\MEMO \date }
\def\memit@m#1{\smallskip \hangafter=0 \hangindent=1in
    \Textindent{\caps #1}}
\def\subject{\memit@m{Subject:}}
\def\topic{\memit@m{Topic:}}
\def\from{\memit@m{From:}}
\def\to{\rel@x \ifmmode \rightarrow \else \memit@m{To:}\fi }
\def\memorule{\medskip\hrule height 1pt\bigskip}  % default definitions
\def\memohead{\centerline{\fourteenrm MEMORANDUM}}% see phyzzx.local
\newwrite\labelswrite
\newtoks\rw@toks
\def\letters{\lettersize
   \headline=\letterheadline \footline=\letterfootline
   \immediate\openout\labelswrite=\jobname.lab}

\let\letterhead=\rel@x
\def\addressee#1{\medskip\line{\hskip 0.75\hsize plus\z@ minus 0.25\hsize
                               \the\date \hfil }%
   \vskip \lettertopskip
   \ialign to\hsize{\strut ##\hfil\tabskip 0pt plus \hsize \crcr #1\crcr}
   \writelabel{#1}\medskip \noindent\hskip -\spaceskip \ignorespaces }
\def\rwl@begin#1\cr{\rw@toks={#1\crcr}\rel@x
   \immediate\write\labelswrite{\the\rw@toks}\futurelet\n@xt\rwl@next}
\def\rwl@next{\ifx\n@xt\rwl@end \let\n@xt=\rel@x
      \else \let\n@xt=\rwl@begin \fi \n@xt}
\let\rwl@end=\rel@x
\def\writelabel#1{\immediate\write\labelswrite{\noexpand\labelbegin}
     \rwl@begin #1\cr\rwl@end
     \immediate\write\labelswrite{\noexpand\labelend}}
\newtoks\FromAddress         \FromAddress={}
\newtoks\sendername          \sendername={}
\newbox\FromLabelBox
\newdimen\labelwidth          \labelwidth=6in
\def\makelabels{\afterassignment\Makelabels \sendersname=}
\def\Makelabels{\FRONTPAGE \letterinfo={\hfil } \MakeFromBox
     \immediate\closeout\labelswrite  \input \jobname.lab\vfil\eject}
\let\labelend=\rel@x
\def\labelbegin#1\labelend{\setbox0=\vbox{\ialign{##\hfil\cr #1\crcr}}
     \MakeALabel }
\def\MakeFromBox{\gl@bal\setbox\FromLabelBox=\vbox{\Tenpoint
     \ialign{##\hfil\cr \the\sendername \the\FromAddress \crcr }}}
\def\MakeALabel{\vskip 1pt \hbox{\vrule \vbox{
        \hsize=\labelwidth \hrule\bigskip
        \leftline{\hskip 1\parindent \copy\FromLabelBox}\bigskip
        \centerline{\hfil \box0 } \bigskip \hrule
        }\vrule } \vskip 1pt plus 1fil }
\def\signed#1{\par \nobreak \bigskip \dt@pfalse \begingroup
  \everycr={\noalign{\nobreak
            \ifdt@p\vskip\signatureskip\gl@bal\dt@pfalse\fi }}%
  \tabskip=0.5\hsize plus \z@ minus 0.5\hsize
  \halign to\hsize {\strut ##\hfil\tabskip=\z@ plus 1fil minus \z@\crcr
          \noalign{\gl@bal\dt@ptrue}#1\crcr }%
  \endgroup \bigskip }
\newbox\letterb@x
\def\lettertext{\par \vskip\parskip \unvcopy\letterb@x \par }
\def\multiletter{\setbox\letterb@x=\vbox\bgroup
      \everypar{\vrule height 1\baselineskip depth 0pt width 0pt }
      \singlespace \topskip=\baselineskip }
\def\letterend{\par\egroup}
%
%%%%%%%%%%%%%%%%%%%%%%%%%%%%%%%%%%%%%%%%%%%%%%%%%%%%%%%%%%%%%%%%%%%%%%%
%
%   Here come macros for title pages.
%
\newskip\frontpageskip
\newtoks\Pubnum   \let\pubnum=\Pubnum
\newtoks\Pubtype  \let\pubtype=\Pubtype
\newif\ifp@bblock  \p@bblocktrue
\def\PH@SR@V{\doubl@true
\baselineskip=24.1pt plus 0.2pt minus 0.1pt
   \parskip= 3pt plus 2pt minus 1pt }
\def\PHYSREV{\papers\PhysRevtrue\PH@SR@V}

\def\titlepage{\FRONTPAGE\papers\ifPhysRev\PH@SR@V\fi
   \ifp@bblock\p@bblock \else\hrule height\z@ \rel@x \fi }
\def\nopubblock{\p@bblockfalse}
\def\endpage{\vfil\break}
\frontpageskip=12pt plus .5fil minus 2pt

\Pubtype={}
\Pubnum={}
\def\p@bblock{\begingroup \tabskip=\hsize minus \hsize
   \baselineskip=1.5\ht\strutbox\topspace
% there used to be a -2 right after topspace which
% has been causing troubles
  \baselineskip\halign to\hsize{\strut ##\hfil\tabskip=0pt\crcr
%  \the\Pubnum\crcr\the\date\crcr\the\pubtype\crcr}\endgroup}
% fong's change
  \the\Pubnum\crcr\the\pubtype\crcr}\endgroup}
\def\title#1{\vskip\frontpageskip \titlestyle{#1} \vskip\headskip }
\def\author#1{\vskip\frontpageskip\titlestyle{\twelvecp #1}\nobreak}

\def\address#1{\par\kern 5pt\titlestyle{\twelvepoint\it #1}}
\def\andaddress{\par\kern 5pt \centerline{\sl and} \address}

\def\abstract{\par\dimen@=\prevdepth \hrule height\z@ \prevdepth=\dimen@
%   \vskip\frontpageskip\centerline{\fourteenrm ABSTRACT}\vskip\headskip }
    \vskip\frontpageskip\centerline{\bf ABSTRACT}\vskip\headskip}

%
%
%%%%%%%%%%%%%%%%%%%%%%%%%%%%%%%%%%%%%%%%%%%%%%%%%%%%%%%%%%%%%%%%%%%%%%%%
%   Miscellaneous macros
%

\def\etal{\hbox{\it et al.}}   
\def\\{\rel@x \ifmmode \backslash \else {\tt\char`\\}\fi }
\def\sequentialequations{\rel@x \if\equanumber<0 \else
  \gl@bal\equanumber=-\equanumber \gl@bal\advance\equanumber by -1 \fi }
%

%
%   We will have name, volume number, page (year)
\def\journal#1&#2&#3(#4){\begingroup \let\journal=\dummyj@urnal
    \unskip~\sl #1\unskip~\bf\ignorespaces #2\rm
    \unskip,~\ignorespaces #3
    (\afterassignment\j@ur \count255=#4)\endgroup\ignorespaces }
\def\j@ur{\ifnum\count255<100 \advance\count255 by 1900 \fi
          \number\count255 }
\def\dummyj@urnal{%
    \toks@={Reference foul up: nested \journal macros}%
    \errhelp={Your forgot & or ( ) after the last \journal}%
    \errmessage{\the\toks@ }}

\def\topspace{\hrule height 0pt depth 0pt \vskip}

\def\Buildrel#1\under#2{\mathrel{\mathop{#2}\limits_{#1}}}
\def\becomes#1{\mathchoice{\becomes@\scriptstyle{#1}}
   {\becomes@\scriptstyle{#1}} {\becomes@\scriptscriptstyle{#1}}
   {\becomes@\scriptscriptstyle{#1}}}
\def\becomes@#1#2{\mathrel{\setbox0=\hbox{$\m@th #1{\,#2\,}$}%
        \mathop{\hbox to \wd0 {\rightarrowfill}}\limits_{#2}}}

\def\VEV#1{\left\langle #1\right\rangle}

\let\int=\intop         
\def\lsim{\mathrel{\mathpalette\@versim<}}
\def\gsim{\mathrel{\mathpalette\@versim>}}
\def\@versim#1#2{\vcenter{\offinterlineskip
        \ialign{$\m@th#1\hfil##\hfil$\crcr#2\crcr\sim\crcr } }}
\def\big#1{{\hbox{$\left#1\vbox to 0.85\b@gheight{}\right.\n@space$}}}
\def\Big#1{{\hbox{$\left#1\vbox to 1.15\b@gheight{}\right.\n@space$}}}
\def\bigg#1{{\hbox{$\left#1\vbox to 1.45\b@gheight{}\right.\n@space$}}}
\def\Bigg#1{{\hbox{$\left#1\vbox to 1.75\b@gheight{}\right.\n@space$}}}
\def\){\mskip 2mu\nobreak }
%
% % % % % % % % % % % % % % % % % % % % % % % % % % % % % % % % % % % %
%
%   Finally, some bug fixings.
%
\let\sec@nt=\sec
\def\sec{\rel@x\ifmmode\let\n@xt=\sec@nt\else\let\n@xt\section\fi\n@xt}
\def\obsolete#1{\message{Macro \string #1 is obsolete.}}
\def\firstsec#1{\obsolete\firstsec \section{#1}}
\def\firstsubsec#1{\obsolete\firstsubsec \subsection{#1}}
\def\thispage#1{\obsolete\thispage \gl@bal\pagenumber=#1\frontpagefalse}
\def\thischapter#1{\obsolete\thischapter \gl@bal\chapternumber=#1}
\def\splitout{\obsolete\splitout\rel@x}
\def\prop{\obsolete\prop \propto }
\def\nextequation#1{\obsolete\nextequation \gl@bal\equanumber=#1
   \ifnum\the\equanumber>0 \gl@bal\advance\equanumber by 1 \fi}
\def\BOXITEM{\afterassigment\B@XITEM\setbox0=}
\def\B@XITEM{\par\hangindent\wd0 \noindent\box0 }
%
%
%%%%%%%%%%%%%%%%%%%%%%%%%%%%%%%%%%%%%%%%%%%%%%%%%%%%%%%%%%%%%%%%%%%%%%%%
%   That's about it
%
\def\phyzzx{PHY\setbox0=\hbox{Z}\copy0 \kern-0.5\wd0 \box0 X}
        
\everyjob{\xdef\today{\monthname~\number\day, \number\year}
        \input myphyx.tex }
\message{ by V.K.}
\catcode`\@=12 % at signs are no longer letters
%
%\dump
% VANILLA.STY
% COPYRIGHT (C) 1985 BY MICHAEL SPIVAK
%
% Moodified for use with PHYZZX by Nigel Goldenfeld
%
% version date 4/7/85
\catcode`\@=11
\font\tensmc=cmcsc10
\def\smc{\tensmc}

\def\hcorrection#1{\advance\hoffset by #1 }
\def\vcorrection#1{\advance\voffset by #1 }
\def\wlog#1{}
\newif\iftitle@
%\outer\def\title{\title@true\vglue 24\p@ plus 12\p@ minus 12\p@
%   \bgroup\let\\=\cr\tabskip\centering
%   \halign to \hsize\bgroup\tenbf\hfill\ignorespaces##\unskip\hfill\cr}
%\def\endtitle{\cr\egroup\egroup\vglue 18\p@ plus 12\p@ minus 6\p@}
%\outer\def\author{\iftitle@\vglue -18\p@ plus -12\p@ minus
%-6\p@\fi\vglue
%    12\p@ plus 6\p@ minus 3\p@\bgroup\let\\=\cr\tabskip\centering
%    \halign to \hsize\bgroup\smc\hfill\ignorespaces##\unskip\hfill\cr}
%\def\endauthor{\cr\egroup\egroup\vglue 18\p@ plus 12\p@ minus 6\p@}
\outer\def\heading{\bigbreak\bgroup\let\\=\cr\tabskip\centering
    \halign to \hsize\bgroup\smc\hfill\ignorespaces##\unskip\hfill\cr}
\def\endheading{\cr\egroup\egroup\nobreak\medskip}

\outer\def\endproclaim{\par\ifdim\lastskip<\medskipamount\removelastskip
  \penalty 55 \fi\medskip\rm}
\outer\def\demo#1{\par\ifdim\lastskip<\smallskipamount\removelastskip
    \smallskip\fi\noindent{\smc\ignorespaces#1\unskip:\enspace}\rm
      \ignorespaces}

\hyphenation{man-u-script man-u-scripts ap-pen-dix ap-pen-di-ces}
\hyphenation{data-base data-bases}
\ifx\amstexloaded@\relax\catcode`\@=13
  \endinput\else\let\amstexloaded@=\relax\fi
\newlinechar=`\^^J
\def\eat@#1{}
\def\Space@.{\futurelet\Space@\relax}
\Space@. %
\newhelp\athelp@
{Only certain combinations beginning with @ make sense to me.^^J
Perhaps you wanted \string\@\space for a printed @?^^J
I've ignored the character or group after @.}
\def\futureletnextat@{\futurelet\next\at@}
{\catcode`\@=\active
\lccode`\Z=`\@ \lowercase
{\gdef@{\expandafter\csname futureletnextatZ\endcsname}
\expandafter\gdef\csname atZ\endcsname
   {\ifcat\noexpand\next a\def\next{\csname atZZ\endcsname}\else
   \ifcat\noexpand\next0\def\next{\csname atZZ\endcsname}\else
    \def\next{\csname atZZZ\endcsname}\fi\fi\next}
\expandafter\gdef\csname atZZ\endcsname#1{\expandafter
   \ifx\csname #1Zat\endcsname\relax\def\next
     {\errhelp\expandafter=\csname athelpZ\endcsname
      \errmessage{Invalid use of \string@}}\else
       \def\next{\csname #1Zat\endcsname}\fi\next}
\expandafter\gdef\csname atZZZ\endcsname#1{\errhelp
    \expandafter=\csname athelpZ\endcsname
      \errmessage{Invalid use of \string@}}}}
\def\atdef@#1{\expandafter\def\csname #1@at\endcsname}
\newhelp\defahelp@{If you typed \string\define\space cs instead of
\string\define\string\cs\space^^J
I've substituted an inaccessible control sequence so that your^^J
definition will be completed without mixing me up too badly.^^J
If you typed \string\define{\string\cs} the inaccessible control sequence^^J
was defined to be \string\cs, and the rest of your^^J
definition appears as input.}
\newhelp\defbhelp@{I've ignored your definition, because it might^^J
conflict with other uses that are important to me.}
\def\define{\futurelet\next\define@}
\def\define@{\ifcat\noexpand\next\relax
  \def\next{\define@@}%
  \else\errhelp=\defahelp@
  \errmessage{\string\define\space must be followed by a control
     sequence}\def\next{\def\garbage@}\fi\next}
\def\undefined@{}
\def\preloaded@{}
\def\define@@#1{\ifx#1\relax\errhelp=\defbhelp@
   \errmessage{\string#1\space is already defined}\def\next{\def\garbage@}%
   \else\expandafter\ifx\csname\expandafter\eat@\string
         #1@\endcsname\undefined@\errhelp=\defbhelp@
   \errmessage{\string#1\space can't be defined}\def\next{\def\garbage@}%
   \else\expandafter\ifx\csname\expandafter\eat@\string#1\endcsname\relax
     \def\next{\def#1}\else\errhelp=\defbhelp@
     \errmessage{\string#1\space is already defined}\def\next{\def\garbage@}%
      \fi\fi\fi\next}
\def\famzero{\fam\z@}

\def\cos{\mathop{\famzero cos}\nolimits}
\def\cosh{\mathop{\famzero cosh}\nolimits}

\def\coth{\mathop{\famzero coth}\nolimits}

\def\exp{\mathop{\famzero exp}\nolimits}

\def\lim{\mathop{\famzero lim}}

\def\sec{\mathop{\famzero sec}\nolimits}
\def\sin{\mathop{\famzero sin}\nolimits}
\def\sinh{\mathop{\famzero sinh}\nolimits}

\def\textfont@#1#2{\def#1{\relax\ifmmode
    \errmessage{Use \string#1\space only in text}\else#2\fi}}
%\textfont@\rm\tenrm
%\textfont@\it\tenit
%\textfont@\sl\tensl
%\textfont@\bf\tenbf
%\textfont@\smc\tensmc
\let\ic@=\/
\def\/{\unskip\ic@}
\def\textfonti{\the\textfont1 }
\def\t#1#2{{\edef\next{\the\font}\textfonti\accent"7F \next#1#2}}
\let\B=\=
\let\D=\.
\def~{\unskip\nobreak\ \ignorespaces}
{\catcode`\@=\active
\gdef\@{\char'100 }}
\atdef@-{\leavevmode\futurelet\next\athyph@}
\def\athyph@{\ifx\next-\let\next=\athyph@@
  \else\let\next=\athyph@@@\fi\next}
\def\athyph@@@{\hbox{-}}
\def\athyph@@#1{\futurelet\next\athyph@@@@}
\def\athyph@@@@{\if\next-\def\next##1{\hbox{---}}\else
    \def\next{\hbox{--}}\fi\next}
\def\.{.\spacefactor=\@m}
\atdef@.{\null.}
\atdef@,{\null,}
\atdef@;{\null;}
\atdef@:{\null:}
\atdef@?{\null?}
\atdef@!{\null!}
\def\srdr@{\thinspace}
\def\drsr@{\kern.02778em}
\def\sldl@{\kern.02778em}
\def\dlsl@{\thinspace}
\atdef@"{\unskip\futurelet\next\atqq@}
\def\atqq@{\ifx\next\Space@\def\next. {\atqq@@}\else
         \def\next.{\atqq@@}\fi\next.}
\def\atqq@@{\futurelet\next\atqq@@@}
\def\atqq@@@{\ifx\next`\def\next`{\atqql@}\else\def\next'{\atqqr@}\fi\next}
\def\atqql@{\futurelet\next\atqql@@}
\def\atqql@@{\ifx\next`\def\next`{\sldl@``}\else\def\next{\dlsl@`}\fi\next}
\def\atqqr@{\futurelet\next\atqqr@@}
\def\atqqr@@{\ifx\next'\def\next'{\srdr@''}\else\def\next{\drsr@'}\fi\next}

\def\textfontii{\the\textfont2 }
\def\{{\relax\ifmmode\lbrace\else
    {\textfontii f}\spacefactor=\@m\fi}
\def\}{\relax\ifmmode\rbrace\else
    \let\@sf=\empty\ifhmode\edef\@sf{\spacefactor=\the\spacefactor}\fi
      {\textfontii g}\@sf\relax\fi}
\def\nonhmodeerr@#1{\errmessage
     {\string#1\space allowed only within text}}
\def\linebreak{\relax\ifhmode\unskip\break\else
    \nonhmodeerr@\linebreak\fi}
\def\allowlinebreak{\relax
   \ifhmode\allowbreak\else\nonhmodeerr@\allowlinebreak\fi}
\newskip\saveskip@
\def\nolinebreak{\relax\ifhmode\saveskip@=\lastskip\unskip
  \nobreak\ifdim\saveskip@>\z@\hskip\saveskip@\fi
   \else\nonhmodeerr@\nolinebreak\fi}
\def\newline{\relax\ifhmode\null\hfil\break
    \else\nonhmodeerr@\newline\fi}
\def\nonmathaerr@#1{\errmessage
     {\string#1\space is not allowed in display math mode}}
\def\nonmathberr@#1{\errmessage{\string#1\space is allowed only in math mode}}
\def\mathbreak{\relax\ifmmode\ifinner\break\else
   \nonmathaerr@\mathbreak\fi\else\nonmathberr@\mathbreak\fi}
\def\nomathbreak{\relax\ifmmode\ifinner\nobreak\else
    \nonmathaerr@\nomathbreak\fi\else\nonmathberr@\nomathbreak\fi}
\def\allowmathbreak{\relax\ifmmode\ifinner\allowbreak\else
     \nonmathaerr@\allowmathbreak\fi\else\nonmathberr@\allowmathbreak\fi}
\def\pagebreak{\relax\ifmmode
   \ifinner\errmessage{\string\pagebreak\space
     not allowed in non-display math mode}\else\postdisplaypenalty-\@M\fi
   \else\ifvmode\penalty-\@M\else\edef\spacefactor@
       {\spacefactor=\the\spacefactor}\vadjust{\penalty-\@M}\spacefactor@
        \relax\fi\fi}
\def\nopagebreak{\relax\ifmmode
     \ifinner\errmessage{\string\nopagebreak\space
    not allowed in non-display math mode}\else\postdisplaypenalty\@M\fi
    \else\ifvmode\nobreak\else\edef\spacefactor@
        {\spacefactor=\the\spacefactor}\vadjust{\penalty\@M}\spacefactor@
         \relax\fi\fi}
\def\newpage{\relax\ifvmode\vfill\penalty-\@M\else\nonvmodeerr@\newpage\fi}
\def\nonvmodeerr@#1{\errmessage
    {\string#1\space is allowed only between paragraphs}}
\def\smallpagebreak{\relax\ifvmode\smallbreak
      \else\nonvmodeerr@\smallpagebreak\fi}
\def\medpagebreak{\relax\ifvmode\medbreak
       \else\nonvmodeerr@\medpagebreak\fi}
\def\bigpagebreak{\relax\ifvmode\bigbreak
      \else\nonvmodeerr@\bigpagebreak\fi}
\newdimen\captionwidth@
\captionwidth@=\hsize
\advance\captionwidth@ by -1.5in
\def\caption#1{}
\def\topspace#1{\gdef\thespace@{#1}\ifvmode\def\next
    {\futurelet\next\topspace@}\else\def\next{\nonvmodeerr@\topspace}\fi\next}
\def\topspace@{\ifx\next\Space@\def\next. {\futurelet\next\topspace@@}\else
     \def\next.{\futurelet\next\topspace@@}\fi\next.}
\def\topspace@@{\ifx\next\caption\let\next\topspace@@@\else
    \let\next\topspace@@@@\fi\next}
 \def\topspace@@@@{\topinsert\vbox to
       \thespace@{}\endinsert}
\def\topspace@@@\caption#1{\topinsert\vbox to
    \thespace@{}\nobreak
      \smallskip
    \setbox\z@=\hbox{\noindent\ignorespaces#1\unskip}%
   \ifdim\wd\z@>\captionwidth@
   \centerline{\vbox{\hsize=\captionwidth@\noindent\ignorespaces#1\unskip}}%
   \else\centerline{\box\z@}\fi\endinsert}
\def\midspace#1{\gdef\thespace@{#1}\ifvmode\def\next
    {\futurelet\next\midspace@}\else\def\next{\nonvmodeerr@\midspace}\fi\next}
\def\midspace@{\ifx\next\Space@\def\next. {\futurelet\next\midspace@@}\else
     \def\next.{\futurelet\next\midspace@@}\fi\next.}
\def\midspace@@{\ifx\next\caption\let\next\midspace@@@\else
    \let\next\midspace@@@@\fi\next}
 \def\midspace@@@@{\midinsert\vbox to
       \thespace@{}\endinsert}
\def\midspace@@@\caption#1{\midinsert\vbox to
    \thespace@{}\nobreak
      \smallskip
      \setbox\z@=\hbox{\noindent\ignorespaces#1\unskip}%
      \ifdim\wd\z@>\captionwidth@
    \centerline{\vbox{\hsize=\captionwidth@\noindent\ignorespaces#1\unskip}}%
    \else\centerline{\box\z@}\fi\endinsert}
\mathchardef\prime@="0230
\def\prime{{{}\prime@{}}}
\def\prim@s{\prime@\futurelet\next\pr@m@s}

\def\,{\relax\ifmmode\mskip\thinmuskip\else\thinspace\fi}
\def\!{\relax\ifmmode\mskip-\thinmuskip\else\negthinspace\fi}
\def\frac#1#2{{#1\over#2}}

\def\:{\nobreak\hskip.1111em{:}\hskip.3333em plus .0555em\relax}
\def\intic@{\mathchoice{\hskip5\p@}{\hskip4\p@}{\hskip4\p@}{\hskip4\p@}}
\def\negintic@{\mathchoice{\hskip-5\p@}{\hskip-4\p@}{\hskip-4\p@}{\hskip-4\p@}}
\def\intkern@{\mathchoice{\!\!\!}{\!\!}{\!\!}{\!\!}}
\def\intdots@{\mathchoice{\cdots}{{\cdotp}\mkern1.5mu
    {\cdotp}\mkern1.5mu{\cdotp}}{{\cdotp}\mkern1mu{\cdotp}\mkern1mu
      {\cdotp}}{{\cdotp}\mkern1mu{\cdotp}\mkern1mu{\cdotp}}}
\newcount\intno@
\def\iint{\intno@=\tw@\futurelet\next\ints@}
\def\iiint{\intno@=\thr@@\futurelet\next\ints@}
\def\iiiint{\intno@=4 \futurelet\next\ints@}
\def\idotsint{\intno@=\z@\futurelet\next\ints@}
\def\ints@{\findlimits@\ints@@}
\newif\iflimtoken@
\newif\iflimits@
\def\findlimits@{\limtoken@false\limits@false\ifx\next\limits
 \limtoken@true\limits@true\else\ifx\next\nolimits\limtoken@true\limits@false
    \fi\fi}
\def\multintlimits@{\intop\ifnum\intno@=\z@\intdots@
  \else\intkern@\fi
    \ifnum\intno@>\tw@\intop\intkern@\fi
     \ifnum\intno@>\thr@@\intop\intkern@\fi\intop}
\def\multint@{\int\ifnum\intno@=\z@\intdots@\else\intkern@\fi
   \ifnum\intno@>\tw@\int\intkern@\fi
    \ifnum\intno@>\thr@@\int\intkern@\fi\int}
\def\ints@@{\iflimtoken@\def\ints@@@{\iflimits@
   \negintic@\mathop{\intic@\multintlimits@}\limits\else
    \multint@\nolimits\fi\eat@}\else
     \def\ints@@@{\multint@\nolimits}\fi\ints@@@}
\def\Sb{_\bgroup\vspace@
        \baselineskip=\fontdimen10 \scriptfont\tw@
        \advance\baselineskip by \fontdimen12 \scriptfont\tw@
        \lineskip=\thr@@\fontdimen8 \scriptfont\thr@@
        \lineskiplimit=\thr@@\fontdimen8 \scriptfont\thr@@
        \Let@\vbox\bgroup\halign\bgroup \hfil$\scriptstyle
            {##}$\hfil\cr}
\def\endSb{\crcr\egroup\egroup\egroup}
\def\Sp{^\bgroup\vspace@
        \baselineskip=\fontdimen10 \scriptfont\tw@
        \advance\baselineskip by \fontdimen12 \scriptfont\tw@
        \lineskip=\thr@@\fontdimen8 \scriptfont\thr@@
        \lineskiplimit=\thr@@\fontdimen8 \scriptfont\thr@@
        \Let@\vbox\bgroup\halign\bgroup \hfil$\scriptstyle
            {##}$\hfil\cr}
\def\endSp{\crcr\egroup\egroup\egroup}
\def\Let@{\relax\iffalse{\fi\let\\=\cr\iffalse}\fi}
\def\vspace@{\def\vspace##1{\noalign{\vskip##1 }}}
\def\aligned{\,\vcenter\bgroup\vspace@\Let@\openup\jot\m@th\ialign
  \bgroup \strut\hfil$\displaystyle{##}$&$\displaystyle{{}##}$\hfil\crcr}
\def\endaligned{\crcr\egroup\egroup}
\def\matrix{\,\vcenter\bgroup\Let@\vspace@
    \normalbaselines
  \m@th\ialign\bgroup\hfil$##$\hfil&&\quad\hfil$##$\hfil\crcr
    \mathstrut\crcr\noalign{\kern-\baselineskip}}
\def\endmatrix{\crcr\mathstrut\crcr\noalign{\kern-\baselineskip}\egroup
                \egroup\,}
\newtoks\hashtoks@
\hashtoks@={#}
\def\format{\crcr\egroup\iffalse{\fi\ifnum`}=0 \fi\format@}
\def\format@#1\\{\def\preamble@{#1}%
  \def\c{\hfil$\the\hashtoks@$\hfil}%
  \def\r{\hfil$\the\hashtoks@$}%
  \def\l{$\the\hashtoks@$\hfil}%
  \setbox\z@=\hbox{\xdef\Preamble@{\preamble@}}\ifnum`{=0 \fi\iffalse}\fi
   \ialign\bgroup\span\Preamble@\crcr}

\def\cases{\left\{\,\vcenter\bgroup\vspace@
     \normalbaselines\openup\jot\m@th
       \Let@\ialign\bgroup$##$\hfil&\quad$##$\hfil\crcr
      \mathstrut\crcr\noalign{\kern-\baselineskip}}

\newif\iftagsleft@
\tagsleft@true
\def\TagsOnRight{\global\tagsleft@false}
\def\tag#1$${\iftagsleft@\leqno\else\eqno\fi
 \hbox{\def\pagebreak{\global\postdisplaypenalty-\@M}%
 \def\nopagebreak{\global\postdisplaypenalty\@M}\rm(#1\unskip)}%
  $$\postdisplaypenalty\z@\ignorespaces}
\interdisplaylinepenalty=\@M
\def\allowdisplaybreak@{\def\allowdisplaybreak{\noalign{\allowbreak}}}
\def\displaybreak@{\def\displaybreak{\noalign{\break}}}
\def\align#1\endalign{\def\tag{&}\vspace@\allowdisplaybreak@\displaybreak@
  \iftagsleft@\lalign@#1\endalign\else
   \ralign@#1\endalign\fi}
\def\ralign@#1\endalign{\displ@y\Let@\tabskip\centering\halign to\displaywidth
     {\hfil$\displaystyle{##}$\tabskip=\z@&$\displaystyle{{}##}$\hfil
       \tabskip=\centering&\llap{\hbox{(\rm##\unskip)}}\tabskip\z@\crcr
             #1\crcr}}
\def\lalign@#1\endalign{\displ@y\Let@\tabskip\centering\halign to \displaywidth
   {\hfil$\displaystyle{##}$\tabskip=\z@&$\displaystyle{{}##}$\hfil
   \tabskip=\centering&\kern-\displaywidth
        \rlap{\hbox{(\rm##\unskip)}}\tabskip=\displaywidth\crcr
               #1\crcr}}
\def\overrightarrow{\mathpalette\overrightarrow@}
\def\overrightarrow@#1#2{\vbox{\ialign{$##$\cr
    #1{-}\mkern-6mu\cleaders\hbox{$#1\mkern-2mu{-}\mkern-2mu$}\hfill
     \mkern-6mu{\to}\cr
     \noalign{\kern -1\p@\nointerlineskip}
     \hfil#1#2\hfil\cr}}}
\def\overleftarrow{\mathpalette\overleftarrow@}
\def\overleftarrow@#1#2{\vbox{\ialign{$##$\cr
     #1{\leftarrow}\mkern-6mu\cleaders\hbox{$#1\mkern-2mu{-}\mkern-2mu$}\hfill
      \mkern-6mu{-}\cr
     \noalign{\kern -1\p@\nointerlineskip}
     \hfil#1#2\hfil\cr}}}
\def\overleftrightarrow{\mathpalette\overleftrightarrow@}
\def\overleftrightarrow@#1#2{\vbox{\ialign{$##$\cr
     #1{\leftarrow}\mkern-6mu\cleaders\hbox{$#1\mkern-2mu{-}\mkern-2mu$}\hfill
       \mkern-6mu{\to}\cr
    \noalign{\kern -1\p@\nointerlineskip}
      \hfil#1#2\hfil\cr}}}
\def\underrightarrow{\mathpalette\underrightarrow@}
\def\underrightarrow@#1#2{\vtop{\ialign{$##$\cr
    \hfil#1#2\hfil\cr
     \noalign{\kern -1\p@\nointerlineskip}
    #1{-}\mkern-6mu\cleaders\hbox{$#1\mkern-2mu{-}\mkern-2mu$}\hfill
     \mkern-6mu{\to}\cr}}}
\def\underleftarrow{\mathpalette\underleftarrow@}
\def\underleftarrow@#1#2{\vtop{\ialign{$##$\cr
     \hfil#1#2\hfil\cr
     \noalign{\kern -1\p@\nointerlineskip}
     #1{\leftarrow}\mkern-6mu\cleaders\hbox{$#1\mkern-2mu{-}\mkern-2mu$}\hfill
      \mkern-6mu{-}\cr}}}
\def\underleftrightarrow{\mathpalette\underleftrightarrow@}
\def\underleftrightarrow@#1#2{\vtop{\ialign{$##$\cr
      \hfil#1#2\hfil\cr
    \noalign{\kern -1\p@\nointerlineskip}
     #1{\leftarrow}\mkern-6mu\cleaders\hbox{$#1\mkern-2mu{-}\mkern-2mu$}\hfill
       \mkern-6mu{\to}\cr}}}
\def\sqrt#1{\radical"270370 {#1}}
\def\dots{\relax\ifmmode\let\next=\ldots\else\let\next=\tdots@\fi\next}
\def\tdots@{\unskip\ \tdots@@}
\def\tdots@@{\futurelet\next\tdots@@@}
\def\tdots@@@{$\mathinner{\ldotp\ldotp\ldotp}\,
   \ifx\next,$\else
   \ifx\next.\,$\else
   \ifx\next;\,$\else
   \ifx\next:\,$\else
   \ifx\next?\,$\else
   \ifx\next!\,$\else
   $ \fi\fi\fi\fi\fi\fi}
\def\text{\relax\ifmmode\let\next=\text@\else\let\next=\text@@\fi\next}
\def\text@@#1{\hbox{#1}}
\def\text@#1{\mathchoice
 {\hbox{\everymath{\displaystyle}\def\textfonti{\the\textfont1 }%
    \def\textfontii{\the\textfont2 }\textdef@@ T#1}}
 {\hbox{\everymath{\textstyle}\def\textfonti{\the\textfont1 }%
    \def\textfontii{\the\textfont2 }\textdef@@ T#1}}
 {\hbox{\everymath{\scriptstyle}\def\textfonti{\the\scriptfont1 }%
   \def\textfontii{\the\scriptfont2 }\textdef@@ S\rm#1}}
 {\hbox{\everymath{\scriptscriptstyle}\def\textfonti{\the\scriptscriptfont1 }%
   \def\textfontii{\the\scriptscriptfont2 }\textdef@@ s\rm#1}}}
\def\textdef@@#1{\textdef@#1\rm \textdef@#1\bf
   \textdef@#1\sl \textdef@#1\it}

\def\textdef@#1#2{\def\next{\csname\expandafter\eat@\string#2fam\endcsname}%
\if S#1\edef#2{\the\scriptfont\next\relax}%
 \else\if s#1\edef#2{\the\scriptscriptfont\next\relax}%
 \else\edef#2{\the\textfont\next\relax}\fi\fi}
\scriptfont\itfam=\tenit \scriptscriptfont\itfam=\tenit
\scriptfont\slfam=\tensl \scriptscriptfont\slfam=\tensl
\mathcode`\0="0030
\mathcode`\1="0031
\mathcode`\2="0032
\mathcode`\3="0033
\mathcode`\4="0034
\mathcode`\5="0035
\mathcode`\6="0036
\mathcode`\7="0037
\mathcode`\8="0038
\mathcode`\9="0039
\def\Cal{\relax\ifmmode\let\next=\Cal@\else
     \def\next{\errmessage{Use \string\Cal\space only in math mode}}\fi\next}
\def\Cal@#1{{\fam2 #1}}
\def\bold{\relax\ifmmode\let\next=\bold@\else
   \def\next{\errmessage{Use \string\bold\space only in math
      mode}}\fi\next}\def\bold@#1{{\fam\bffam #1}}
\mathchardef\Gamma="0000
\mathchardef\Delta="0001
\mathchardef\Theta="0002
\mathchardef\Lambda="0003
\mathchardef\Xi="0004
\mathchardef\Pi="0005
\mathchardef\Sigma="0006
\mathchardef\Upsilon="0007
\mathchardef\Phi="0008
\mathchardef\Psi="0009
\mathchardef\Omega="000A
\mathchardef\varGamma="0100
\mathchardef\varDelta="0101
\mathchardef\varTheta="0102
\mathchardef\varLambda="0103
\mathchardef\varXi="0104
\mathchardef\varPi="0105
\mathchardef\varSigma="0106
\mathchardef\varUpsilon="0107
\mathchardef\varPhi="0108
\mathchardef\varPsi="0109
\mathchardef\varOmega="010A
\font\dummyft@=dummy
\fontdimen1 \dummyft@=\z@
\fontdimen2 \dummyft@=\z@
\fontdimen3 \dummyft@=\z@
\fontdimen4 \dummyft@=\z@
\fontdimen5 \dummyft@=\z@
\fontdimen6 \dummyft@=\z@
\fontdimen7 \dummyft@=\z@
\fontdimen8 \dummyft@=\z@
\fontdimen9 \dummyft@=\z@
\fontdimen10 \dummyft@=\z@
\fontdimen11 \dummyft@=\z@
\fontdimen12 \dummyft@=\z@
\fontdimen13 \dummyft@=\z@
\fontdimen14 \dummyft@=\z@
\fontdimen15 \dummyft@=\z@
\fontdimen16 \dummyft@=\z@
\fontdimen17 \dummyft@=\z@
\fontdimen18 \dummyft@=\z@
\fontdimen19 \dummyft@=\z@
\fontdimen20 \dummyft@=\z@
\fontdimen21 \dummyft@=\z@
\fontdimen22 \dummyft@=\z@
\def\fontlist@{\\{\tenrm}\\{\sevenrm}\\{\fiverm}\\{\teni}\\{\seveni}%
 \\{\fivei}\\{\tensy}\\{\sevensy}\\{\fivesy}\\{\tenex}\\{\tenbf}\\{\sevenbf}%
 \\{\fivebf}\\{\tensl}\\{\tenit}\\{\tensmc}}
\def\dodummy@{{\def\\##1{\global\let##1=\dummyft@}\fontlist@}}
\newif\ifsyntax@
\newcount\countxviii@
\def\newtoks@{\alloc@5\toks\toksdef\@cclvi}
\def\nopages@{\output={\setbox\z@=\box\@cclv \deadcycles=\z@}\newtoks@\output}
\def\syntax{\syntax@true\dodummy@\countxviii@=\count18
\loop \ifnum\countxviii@ > \z@ \textfont\countxviii@=\dummyft@
   \scriptfont\countxviii@=\dummyft@ \scriptscriptfont\countxviii@=\dummyft@
     \advance\countxviii@ by-\@ne\repeat
\dummyft@\tracinglostchars=\z@
  \nopages@\frenchspacing\hbadness=\@M}
\def\wlog#1{\immediate\write-1{#1}}
\catcode`\@=\active

\font\bfgreek=cmmib10 scaled\magstep1

\hcorrection{1truein}
\vcorrection{1 truein}

% For use with PHYZZX

\def\bmath{\fam1\bfgreek\textfont1=\bfgreek}

\def\rf#1{$^{#1}$}

\def\ltwid{\raise.3ex\hbox{$<$\kern-.75em\lower1ex\hbox{$\sim$}}}
\def\gl{\raise.5ex\hbox{$>$}\kern-.8em\lower.5ex\hbox{$<$}}
\def\gtwid{\raise.3ex\hbox{$>$\kern-.75em\lower1ex\hbox{$\sim$}}}

\def\3he{$^3\text{He}$}
\def\4he{$^4\text{He}$}

\def\etal{{\it et al.}}

  %These three spacing deals will eventually
    % be fixed by Tony.

%\font\sit=amti9
%\outer\def\Address{\bgroup\let\\=\cr\tabskip\centering
%    \halign to \hsize\bgroup\sit\hfill\ignorespaces##\unskip\hfill\cr}
%\def\Endaddress{\cr\egroup\egroup\nobreak\medskip}

\def\VEC#1{\bold{#1}}
%\newcount\secnum
%\newcount\eknum
\newcount\refnum
\global\refnum=0
%\global\eknum=0
%\global\secnum=0
%\def\rf{\global\advance\refnum by 1 $^{\the\refnum\text{ }}$}
\def\bib{\global\advance\refnum by 1\item{\the\refnum.}}
%\def\e1{\global\advance\eknum by 1}
%\def\etag{\global\advance\eknum by 1\tag \the\eknum}
%\def\setag{\global\advance\eknum by 1\tag \the\secnum.\the\eknum}
%\outer\def\section{\vfill\eject\global\advance\secnum by 1\eknum=0
%\bgroup\let\\=\cr\tabskip\centering
%    \halign to
%\hsize
%\bgroup\bigfonts\hfill\ignorespaces##\unskip\hfill\cr}
%\def\endsection{\cr\cr\egroup\egroup\nobreak}
\def\a0{\text{ \AA}}
%\def\Refs{\vfill\eject\centerline{\bigfonts
%References}\global\refnum=0\bigskip}
%\def\Figs{\vfill\eject\global\refnum=0\cl{\bigfonts Figure
%Captions}\bigskip}

%\outer\def\Title{
%\bgroup\baselineskip 20pt\let\\=\cr\tabskip\centering
%    \halign to
%\hsize
%\bgroup\bigfonts\hfill\ignorespaces##\unskip\hfill\cr}
%\def\Endtitle{\cr\cr\egroup\egroup\nobreak}

\def\lda{\lambda}

\def\hz{\hat \zeta}

\def\pa{\partial}
\def\vecr{\VEC r}

\def\rt  {({\VEC r},t)}
\def\sqr#1#2{{\vcenter{\vbox{\hrule height.#2pt
     \hbox{\vrule width.#2pt height#1pt \kern#1pt
     \vrule width.#2pt}
     \hrule height.#2pt}}}}
 
\def\bm#1{\hbox{$\bmath #1$}}

\def\vecsigma{\bm\sigma}
\def\hm{\hat M}
\HOFFSET=0.375truein

%\PHYSREV
%\unnumberedchapters
%\def\vs{{\it versus}}
\tolerance 2000
%\nopubblock
\pubnum={NSF-ITP-93-137}
\date{}
\pubtype{}

\titlepage
\title{\bf Stability and Kinetics of Step Motion on
      Crystal Surfaces}

\author{Fong Liu\rf{1} and Horia Metiu\rf{2}}
\address{\rf1 Institute for Theoretical Physics and Center for
Quantized\break Electronic Structures, University of
California, Santa Barbara, CA 93106}
\medskip
\address{\rf2Department of Chemistry,
University of California,\break Santa Barbara, CA 93106}
\vfil

\abstract
The kinetics of monoatomic steps in diffusion-controlled
crystal growth and evaporation processes are investigated
analytically using a Green's function approach.
Integro-differential equations of motion for the steps are derived;
and a systematic linear stability analysis is carried out treating
simultaneously perturbations both along and perpendicular
to the steps. Morphological fluctuations of steadily moving steps
in response to ambient thermodynamic noises
are also studied within a general Langevin formalism.
Finally, a phase field model is developed to investigate the
time-dependent, collective motion of steps.
An application of the model to a finite step train
recovers a variety of kinetic behaviors such as the
bunching and spreading of steps.

\noindent
\normalspace
%\doublespace

\bigskip\bigskip\noindent
PACS numbers: 61.50.Cj, 05.40.+j, 68.35.Ja
%\medskip\submit{}
\bigskip\vfil
\endpage

\chapter{INTRODUCTION}

 The kinetics of atomic steps play a central role in the growth of
 singular crystal surfaces from vapor, liquid,
 and solid phases, in a wide variety of materials.
 Step motion is also a predominant elementary process governing
 such macroscopic changes of crystal surface topography as
 dissolution, faceting, etching and coarsening.
 Numerous investigations, both theoretical and experimental,
 have been devoted to understand the morphology and dynamics
 of atomic steps. Recent resurgence of interests in
 step dynamics is largely stimulated by
 the development of crystal growth methods
 with atomic-precision, and by the advances
 in high-resolution microscopies.

 This paper addresses several theoretical aspects of the
 kinetics of diffusively coupled steps. Our investigation
 is based upon the classical model of step kinetics developed
 by Burton, Cabrera, and Frank (BCF)\rlap.\Ref\bcf{W.K.
 Burton, N. Cabrera, and F.C. Frank, \journal Phil. Trans. R. Soc.
 A&243&299(1951).}  The ideal BCF model visualizes the
 singular crystal surface as consisting of closely-packed
 terraces separated by elementary steps. Growth of crystal
 then proceeds through the lateral progression of these steps by
 incorporating diffusing adatoms from the terraces.
 The BCF model, supplemented with later modifications and
 extensions, has been extensively investigated and
 remains the most mathematically well-posed model of crystal
 growth by step flow dynamics.
 Although we restrict ourselves to the behavior
 of monoatomic steps governed by surface diffusion,
 and to situations where other assumptions of the BCF model are
 valid, we hope that this study can stimulate progress in elucidating
 the role of steps (not necessarily elementary) under other
 growth conditions, e.g., the growth of crystal by molecular
 epitaxy, or the ledge growth by volume diffusion in solid/solid phase
 transformations.

 One central question in step kinetics concerns the
 morphological stability of step patterns.
 Steps are seldom seen to be equally spaced. It is also
 observed that fluctuations in the distance
 between steps can often lead to the bunching of them
 into multi-step bands. Furthermore, meandering of the steps
 along the step direction on scales much greater than molecular
 lengthscale is commonplace. Bales and Zangwill\Ref\bales{
 G.S. Bales and A. Zangwill, \journal Phys. Rev. B.
 &41&5500(90); \journal ibid. &48&2024E(93).}
 recently pointed out that, in addition to the effect of
 equilibrium fluctuations, a diffusive instability can
 contribute to the meandering of steps under growth situations.
 To address the stability issue, linear stability analyses,
 treating separately fluctuations in the spacing between
 adjacent steps and that along the step direction, have been
 carried out. Nevertheless, two popular approximations are
 often adopted.  First,  when considering perturbations in the step
 spacings,  it is commonly assumed that
 the velocity of a step only depends upon the width of its
 two adjacent terraces. Second, in treating fluctuations along the
 steps, and in studying the motion of steps in general, a
 quasi-static approximation is used by
 assuming sufficient slow motion of steps.
 While these approximations are believed to be good
 for widely separated steps and under small driving forces,
 a fully time-dependent analysis is warranted in the opposite
 situations.
 As emphasized by Ghez and Iyer\rlap,\Ref\ghez{
 R. Ghez and S.S. Iyer, \journal IBM J. Res. Dev. &32&804(88).}
 such situations may arise particularly in certain epitaxy
 experiments where the so-called
 fast steps are encountered and where nearby steps are strongly
 coupled by their overlapping diffusion fields.
 Linear stability analysis which takes account of the overlap
 of diffusion fields has recently been carried out
 numerically\rlap.\REFS\gheztwo{R. Ghez, H.G. Cohen, and J.B. Keller,
 \journal App. Phys. Lett. &56&1977(90).}\REFSCON\vlachos{
 D.G. Vlachos and K.F. Jensen, \journal Surf. Sci. &262&359(1992).}
 \refsend
 In this paper, the step flow problem is formulated using a
 Green's function approach. Time-dependent integro-differential
 equations of motion for the step positions are derived, from
 which a systematic linear stability analysis is carried out.
 This formulation allows a simultaneous treatment of generic
 perturbations both along step extension and in the step
 spacings. The validity of the quasi-static approximation and
 the effect of the overlap of terrace diffusion fields are
 also examined. These works are discussed in section 2
 through section 4. Section 5 analyzes the effect of
 asymmetric attachment kinetics on the linear stability
 of infinite step trains, under both growth and evaporation
 situations.

 In section 6, we discuss the morphological fluctuations of
 steps during non-equilibrium growth using a Langevin
 formalism. In this formalism, local thermodynamical
 noises on the terraces and on the steps are assumed
 to be the sole sources of fluctuations which get amplified
 dynamically during growth and lead to the roughness of steps.
 We show that this method consistently recovers the
 equilibrium fluctuations. Whilst step roughness in
 equilibrium does not depend on step spacing, we found that
 the diffusive coupling of steps in motion during growth
 gives rise to a correlation between fluctuations on
 different steps.

 While stability analysis gives important information with regard
 to the role of step kinetics in crystal growth, more
 insights can only be provided by detailed investigations of
 the temporal development of step configurations.
 It is crucially required, for example, to interpret in-situ
 experimental observations. In section 7, we develop
 a phase field approach to study the time-dependent
 motion of finite and generically non-equidistant step trains.
 This method is numerically efficient, easily adaptable to
 two-dimensions, compatible with shape fluctuations along the
 step extension, and able to simulate complicated
 time-dependent morphology of step patterns.

\chapter{THE BCF MODEL}

Consider the mathematical formulation of the original
BCF model, for a close-packed crystal surface below its
roughening transition temperature, growing from
a supersaturated mother medium.
We concentrate on the layer-by-layer growth mechanism
characterized by the diffusion-controlled propagation
of monoatomic steps, and neglect entirely the effect of
two-dimensional island nucleation, under the assumption
of small supersaturations. For concreteness, it is
further assumed that the growth units are incorporated to
the steps through diffusion along the terraces only.
Although this mode of building growth units is
predominant in vapor and solution growth, the alternative mode
of depositing growth units directly from
the bulk mother medium is important in certain
solution growth environments and particularly in solid-solid
phase transformations.
The BCF formulation is equally applicable to crystal evaporation
and dissolution processes.

In the BCF model, transport of adatoms on the terraces
proceeds through surface diffusion according to
 $$ \frac{\pa c}{\pa t} =D\nabla^2 c - \frac{c-c_{\infty}
   }{\tau_s}   \eqn\diff $$
where $c\rt$ is the adatom density,  $\tau_s$ is the mean
lifetime of adatom on the
terraces before evaporating into the vapor,
$c_{\infty}$ is the adatom density far away from the steps,
and $D$ is a diffusion coefficient
assumed isotropic over all terraces.
In the present macroscopic description, steps are regarded
as mathematically sharp dividing lines between terraces.
A crucial assumption, inherent in most continuum treatments,
demands that the monoatomic steps are molecularly rough
with numerous kinks excitations so as to behave as ideal line sinks
for growth units.
The condition of molecular roughness of steps is also a
prerequisite for us to safely neglect lateral diffusion of
adatoms along the steps, and to assume isotropic advancement of steps.
A lower kink density hence smoother steps often implies stronger
effect of attachement kinetics and more pronounced anisotropy in
step motion and morphology.
These and other assumptions of the BCF approach must be carefully
examined when applying the original model to experiments,
particularly to such nonconventional situations as
molecular beam epitaxy\refmark{\ghez} at low temperatures.

Relevant microscopic details of the atomic processes
near the steps are reflected in the boundary conditions on
the sharp steps. Firstly, conservation of materials requires that
the normal velocity of a step satisfies
 $$  c_r v_n = D \left[ (\nabla c)_+ - (\nabla c)_-
    \right]\cdot {\VEC n}   \eqn\latent $$
where $c_r$ is the change of atomic density in the area
swept by the moving step,
${\VEC n}$ is the unit normal to the step, and $(\nabla c)_{\pm}$
are gradients computed on the two sides of a
step, respectively. Secondly, another boundary condition
relates the density of adatoms at the step to its local
equilibrium value, taking into account the curvature
correction via the Gibbs-Thompson relation:
 $$ c_{step}=c_{0}\left(1+ \frac{\gamma \Omega}{k_BT}
    \kappa\right) \eqn\bound $$
where $c_{0}$ is the adatom density for a straight step
in equilibrium, $\gamma$ is the isotropic line tension of
the step, $\Omega$ is the atomic area of the solid,
and $\kappa$ is the step curvature.
The density of adatoms far away from the steps is maintained
above the equilibrium density at $c_{\infty}>c_{0}$ through
supersaturation, providing the driving force for step propagation.
We have neglected the effect of crystalline anisotropy
and also temporarily ignored the elevation of adatom
density at the steps due to attachment kinetics.
Equations \diff\ through \bound\
complete the description of the {\it symmetric} BCF model for
crystal growth by the step flow mechanism, where
atomic processes on both sides of a step contribute
equivalently to step progression.
The corresponding {\it asymmetric} version of the model,
accounting for nonequivalent attachment kinetics as first discussed
by Schwoebel and Shipsey\rlap,\Ref\schwoebel{R.L. Schwoebel
and E.J. Shipsey, \journal J. App. Phys. &37&3682(66); R.L.
Schwoebel, \journal J. App. Phys.  &40&614(69).}
will be considered in section 5.

Multiple lengthscales are involved in the current problem,
whose relative ratio characterize different growth regimes and
determine the validity of various approximations.
With the assumption of mathematically sharp steps, the continuum
model is understood to apply on length scales much longer than the
microscopic capillary length $d_0$ (to be defined below).
Among the relevant macroscopic lengthscales,
the mean adatom diffusion length on the terraces is given by
$x_s\equiv \sqrt{D\tau_s}$, and the typical step spacing
is denoted by $\lambda$.
Another quantity $l=2D/v$ with dimensionality of length
can be conveniently introduced as an alternative measure of
the speed of step motion. Defining a quantity
 $$ u\rt = \frac{c_{\infty}-c\rt}{c_r}   \eqn\defofu $$
and measuring time in units of $\tau_s$, length in units
of $x_s$, and velocity in units of $\sqrt{D/\tau_s}$,
the equations of motion can be written in the dimensionless
form
 $$ \frac{\pa u}{\pa t}= \nabla^2 u - u
  \eqn\dimone $$
 $$ v_n= - \left[ (\nabla u)_+ - (\nabla u)_-
    \right]\cdot {\VEC n}  \eqn\dimtwo $$
 $$ u_{step}= \Delta - d_0 \kappa  \eqn\dimthree $$
where $\Delta=(c_{\infty}-c_{0})/c_r$ is the dimensionless
supersaturation and $d_0=\gamma\Omega c_{0} /(k_BT c_r x_s)$
is the dimensionless capillary length.
The problem defined by \dimone\ through \dimthree\
resembles closely the problem of
solidification\rlap,\Ref\nigel{J.S. Langer, \journal Rev. Mod.
Phys. &50&1(80); N.D. Goldenfeld, in {\sl Metastability and
Incompletely Posed Problems}, edited by S. Antman et al.
(Springer-Verlag, Berlin, 1987); D.A. Kessler, J. Koplik,
and H. Levine, \journal Adv. Phys. &34&123(90).}
 in spite of
the extra homogeneous term $-u$ on the r.h.s. of \dimone.

\chapter{EQUATION OF MOTION FOR THE STEPS}

The moving boundary problem \dimone\ to \dimthree\
belong to a general category of problems known as
the Stefan problem, and is strongly non-linear in character.
One must solve a diffusion-like equation subject to
boundary conditions which themselves
depend upon the solution of the same diffusion equation.
Complete analytic treatment of the entire problem in
2-space and 1-time dimensions is difficult.
Even the linear stability analysis has only been carried out
for synchronously moving equidistant steps, under
quasi-static approximations. Studies of time-dependent step
motion, particularly for non-equidistant steps, are
even more limited.
Fortunately, progress can be made analytically by reformulating
the full 2+1 dimensional problem in terms of a closed set of
integro-differential equations for the positions of the steps.
This boundary integral method, using Green's functions,
was first put forward by Langer and Turski in their study of
directional solidification fronts\rlap.\Ref\langer{J.S. Langer
and L.A. Turski, \journal Acta Metall. &25&1113(77);
J.S. Langer, \journal Acta Metall. &25&1121(77).}
A straightforward application of their procedure
to our present problem yields the
dynamical evolution equations for the moving steps,
starting from which many detailed analyses can be made.

Consider an infinite train of non-overlapping steps on the
crystal surface shown schematically in Fig.1.
\FIG\figone{Schematic representation of a train of steps
moving in the $z$-direction.}
The time-dependent position of the $n$th step is parameterized by
$z=\zeta^n(x,t)$. In a coordinate frame moving with
velocity $V$ in the
positive $z$-direction, the free boundary problem takes the form:
 $$ \left( \nabla^2 + V\frac{\pa}{\pa z}
-1-\frac{\pa}{\pa t}\right)u(x,z,t)=0
\eqn\movfr  $$
 $$\left(V+\frac{d\zeta^n}{dt}\right)dx = -\left[ (\nabla u)^n_+
    -(\nabla u)^n_{-} \right]\cdot d\vecsigma^n
\eqn\eimtwo $$
 $$ u(x,\zeta^n,t)= \Delta -d_0 \kappa\{\zeta^n\}  \eqn\eimthree $$
where the superscripts label the steps and the vector
step length-element $d\vecsigma$ is defined according to Fig.1.
To proceed, we define a Green's function in the moving frame
by
  $$ \left( \nabla_1^2 - V\frac{\pa}{\pa z_1}
-1+\frac{\pa}{\pa t_1}\right)G(p|p_1)=
   -\delta(p-p_1)  \eqn\green  $$
with the short notation $p=(x,z,t)$, $p_1=(x_1,z_1,t_1)$.
The causality condition that $G(p|p_1)=0$ for $t<t_1$ is
also imposed. For an arbitrary space-time
point $p$, we have from \movfr\ and \green,
 $$\eqalign{
 u(p)= &  \int_{-\infty}^{t} dt_1\int_{S_1}
     d\vecsigma_1\cdot
   \left[u(p_1)\nabla_1G(p|p_1)-G(p|p_1)\nabla_1u(p_1)\right]
    \cr
   & +V\int_{-\infty}^{t} dt_1\int_{\Lambda_1}
  dx_1dz_1 \frac{\pa}{\pa z_1} \left[u(p_1)G(p|p_1)\right]
    }   \eqn\symfull $$
where the Green's theorem has been applied to the region $\Lambda_1$
containing the point $p$ but excluding all the steps bounded
by the boundary surface $S_1$. The volume contribution
in \symfull\ vanishes identically because the
integration over $dz_1$ gives cancelling terms from
two sides of each step. This cancellation is
specific to the symmetric model, where
the continuity of adatom density across each step is maintained.
Further application of this continuity property to the
first part of the surface integral in \symfull\ also gives
a null result.  Hence \symfull\ reduces to
 $$  u(p)=\sum_{m=-\infty}^{+\infty} \int_{-\infty}^{t} dt_1
    \int_{p_1\in m} G(p|p_1)\left[
       (\nabla_1u(p_1))_- -(\nabla_1u(p_1))_+\right]
     \cdot d\vecsigma_{1+}   \eqno\eq $$
which, upon using the boundary condition \eimtwo\
and letting $p$ approach the $n$th step, leads to
the equation of motion for the $n$th step:
 $$ \Delta- d_0\kappa\{\zeta^n\}
  = \sum_{m=-\infty}^{+\infty}
  \int_{-\infty}^{t} dt_1 \int_{-\infty}^{+\infty}
  dx_1 G\left(x, \zeta^n,t|x_1,\zeta^m_1,t_1
   \right)\left(V+ \frac{d\zeta^m_1}{dt_1}\right)  .
\eqn\symfullb   $$
Here the curvature is by definition
 $$ \kappa\{\zeta^n\}=-\frac{d^2\zeta^n}{dx^2}\left[
    1+\left(\frac{d\zeta^n}{dx}\right)^2
    \right]^{-3/2} .  \eqno\eq $$
This set of closed, nonlinear, integro-differential equations
describe the time-dependent evolution of an infinite
train of steps mutually coupled through the overlapping
adatom diffusion fields. With little modification, the
boundary integral approach can also be used to investigate
pattern selection problems for spiral steps.
Finally, we point out that, in the present consideration, no energetic
interaction between steps, such as that due to elasticity,
is yet incorporated.

\chapter{LINEAR STABILITY ANALYSIS}

The boundary integral formulation makes possible certain
analytical treatments otherwise infeasible using the
original full diffusion model. For example, this approach
is particularly suitable for determining various steady state
solutions and for analyzing their stabilities.
Although equation \symfullb\ supports in general steady state
solutions of both straight and curved steps, we restrict our
consideration in this paper to straight steps only.

This section presents a linear stability analysis of
uniformly moving step trains.
The advantages of using \symfullb\ for linear stability
analysis, rather than using the full diffusion equation,
are two-fold.
First the method is systematic and allows us
to treat simultaneously perturbations both along and
perpendicular to the steps. Second, the calculation
can be carried out without having to use the
quasi-static approximation, which becomes unreliable for
both fast-moving and densely-packed step trains.
The situation of fast step trains is likely to
occur in MBE growth where the dimensionless supersaturation
is typically large\rlap.\refmark{\ghez}
Some of our results reported
here were previously known but were obtained using
different methods under quasi-static approximations.

First we concentrate on steady state solutions
for straight steps. Consider an infinite train of
equidistant steps moving uniformly with velocity $V$
in the positive $z$-direction.
In the co-moving frame of the steps,
letting $\zeta^n(x,t)=n\lambda$
with $\lambda$ being the step spacing, and using the
real-space representation of the Green's function
 $$ G(p|p_1)=
 \frac{e^{-(t-t_1)}}{4\pi (t-t_1)}\exp \left[
 -\frac{(x-x_1)^2+(z-z_1+V(t-t_1))^2}{4(t-t_1)}\right], \eqno\eq $$
we obtain from \symfullb,
 $$ \Delta= V\sum_{m=-\infty}^{+\infty}
       \int_{0}^{+\infty}d\tau\int_{-\infty}^{+\infty}
       dx_1 G(0,0,\tau|x_1,m\lambda,0),   \eqno\eq $$
which leads to the transcendental equation\refmark{\ghez}
 $$ 2\Delta = \left(\frac{\alpha_+ -\alpha_-}
    {\alpha_+ +\alpha_-}\right) \left[
    \coth  \frac{\alpha_+\lambda}{2}+\coth \frac{\alpha_-\lambda}{2}
    \right]   \eqn\stead $$
where we have defined for convenience $$ \alpha_{\pm}
\equiv \sqrt{1+\frac{1}{l^2}} \pm\frac{1}{l} .  \eqno\eq $$
Equation \stead\  can be solved numerically for the velocity of steps
$V=V(\Delta, \lambda)$, which is monotonically increasing
in both $\Delta$ and $\lambda$.

Next let us analyze the morphological stability
of this system of equidistant steps.
Consider small-amplitude perturbations to the steady state
step profiles in the general form
 $$ \hz^n(x,t)= \zeta^n(x,t)-n\lambda   \eqno\eq $$
for all $n$. Equations governing the time evolution
of perturbations are derived by expanding \symfullb\ in
powers of the small quantity $\hz^n$.  We shall use
the Green's function in a Fourier representation
 $$ G(p|p_1)=\int\frac{d\omega dk}{(2\pi)^2}
      e^{ik(x-x_1)+i\omega (t-t_1)}
     \frac{
  e^{-(\zeta^n-\zeta^m_1)/l-|\zeta^n-\zeta^m_1|M} }{2M(k,\omega)}
  \eqno\eq $$
where
 $$ M(k,\omega) \equiv \sqrt{k^2+1+\frac{1}{l^2}+i\omega } \eqno\eq $$
with a positive real part.
The integrand in the Green's function can be readily expanded
to first order in perturbation to read
 $$ e^{-(\zeta^n-\zeta^m_1)/l-
  |\zeta^n-\zeta^m_1|M}=I^{(0)}_{nm}+I^{(1)}_{nm}+... , \eqno\eq $$
where the zeroth order term is given by
 $$ I^{(0)}_{nm}=e^{-(n-m)\lambda/l-
  |n-m|\lambda M}  \eqno\eq $$
and the first order term takes the form
 $$ I^{(1)}_{nm}= \left\{
  \eqalign{
 & -(\hz^n-\hz^n_1)/l-|\hz^n-\hz^n_1|M
    \qquad \hbox{for $n=m$} \cr
   & -(Ml+1)e^{-(M+1/l)(n-m)\lambda}(\hz^n-\hz^m_1)/l \qquad \hbox{
   for $n>m$}  \cr
   & (Ml-1)e^{(M-1/l)(n-m)\lambda}(\hz^n-\hz^m_1)/l \qquad \hbox{
   for $n<m$} .  }\right.   \eqno\eq $$
Now these expressions can then be inserted into the r.h.s. of \symfullb\
which has the expansion
 $$ \int dt_1dx_1\frac{d\omega dk}{(2\pi)^2}
 \frac{e^{ik(x-x_1)+i\omega (t-t_1)}}{2M}
 \sum_m\left[ \frac{d\hz^m_1}{dt_1}I^{(0)}_{nm}+VI^{(1)}_{nm}
 \right].  \eqno\eq $$
It is more convenient to work in the Fourier space
by defining Fourier transforms as, for example,
 $$ \hz^n(k,\omega)= \int_{-\infty}^{+\infty}
   dt \int dx e^{-ikx-i\omega t}\hz^n(x,t) . \eqno\eq    $$
After grouping together all the terms that are
linear in perturbation in the expansion of equation \symfullb,
we obtain the linear stability equation
 $$ {\Cal L}(k,\omega,\hz)=0   \eqn\stab $$
where
 $$\eqalign{
  {\Cal L}(k,\omega,\hz) &=
  \left[ \frac{i\omega l^2}{2}+1 -Ml\left(\Delta-d_0l k^2
 +\frac{1}{e^{\alpha_+\lambda}-1}
 -\frac{1}{e^{\alpha_-\lambda}-1}
  \right) \right]
  \hz^n(k,\omega)   \cr
 & +\left(\frac{i\omega l^2}{2}+1+Ml \right)
  \sum_{m<n} e^{-(M+1/l)(n-m)\lambda} \hz^m(k,\omega) \cr
 & +\left(\frac{i\omega l^2}{2}+1-Ml \right)
  \sum_{m>n} e^{-(M-1/l)(m-n)\lambda}
 \hz^m(k,\omega) .  } \eqn\stabtwo $$
Linear dispersion relations under general perturbations are
solutions of \stab, as we shall show below.

Before moving on, we reiterate that, although we have so far
considered only straight steps, equation \symfullb\ also
has steady state solutions corresponding to curved steps.
In general, these solutions can be
approximately evaluated, in the small amplitude
limit, by expanding \symfullb\ to higher orders in $\hz$.
The linear stability of these solutions can in turn be
similarly investigated. However, these calculations are
in practice quite cumbersome\rlap.\refmark{\langer}

\section{Stability of an Isolated Step}
As a special case, let us consider the linear stability
of an isolated step.  This situation is equivalent to the
limit of infinite step spacing $\lambda =\infty$.
Stability equation \stab\ simplifies to
 $$ \frac{i\omega l^2}{2}+1 -\left( \Delta
     -d_0lk^2 \right)lM(k,\omega)=0  \eqno\eq $$
which gives the linear dispersion relation in the closed form
 $$ \frac{i\omega l^2}{2}=\left(\Delta-d_0lk^2\right)
    \sqrt{ l^2k^2+l^2-1+\left(
  \Delta-d_0lk^2\right)^2 }-1+\left(\Delta-d_0lk^2\right)^2
 \eqn\planarspec $$
where the quantity $l=\sqrt{1-\Delta^2}/\Delta$ is
solved from \stead.
This dispersion relation is plotted in Fig.2.
\FIG\figtwo{Linear stability spectrum \planarspec\
of an isolated step against shape perturbations,
in the symmetric model. Curves correspond, from top down,
to values of line tension ($d_0$) $0.25$, $1$, and $2$ times the
critical value $\Delta^2\sqrt{1-\Delta^2}/2$.}

It is evident from Fig.2 that, within certain parameter
range, a straight step becomes linearly unstable against
infinitesimal, long-wavelength fluctuations.
This diffusive instability is of the same nature as that
discovered by Mullins and Sekerka\Ref\mullinssek{W.W. Mullins
and R.F. Sekerka, \journal J. Appl. Phys. &34&323(63);
\journal ibid, &35&444(64).} for planar solidification fronts.
The relevance of Mullins-Sekerka instability in step flow
was first pointed out by Bales and Zangwill\rlap.\refmark{
\bales}
The step problem in question here, although resembling closely
that of solidification, has important differences, in two
respects. First, in solidification, steady state planar fronts
growing into supercooled liquid do not exist at general
supercoolings, in contrast with the
situation here where uniformly moving straight steps are valid
solutions at arbitrary supersaturation. Second,  a planar
front in solidification is always linearly unstable to
perturbations of sufficiently long wavelength, with the effect of
surface tension merely stablizing the front on short lengthscales.
But, again by contrast, straight steps can be completely
stablized by strong enough line tension on all scales (see
below). Both differences between the two problems
are due to the presence of the extra lengthscale,
the adatom diffusion length $x_s$, in the step problem.

It follows from \planarspec\ that, as long as $d_0\ge l\Delta^3/2$,
a straight step is linearly stable against all infinitesimal
perturbations, i.e., $\hbox{Re}(i\omega)\leq 0$ for all $k$.
Since $l=\sqrt{1-\Delta^2}/\Delta$, this condition is equivalent to
 $$d_0\ge \frac{\Delta^2\sqrt{1-\Delta^2}}{2} .  \eqn\stlim $$
On the other hand, for smaller values of
$d_0$, the straight step is unstable against long wavelength
perturbations. The critical wavenumber below which perturbation
grows is given by
 $$ k_c=\sqrt{\frac{\Delta}{d_0l}}\left[
   1-\frac{d_0}{2\Delta^3 l}-\sqrt{
   \frac{d_0}{\Delta^3 l}+
  \frac{d_0^2}{4\Delta^6 l^2}}\right]^{1/2} . \eqno\eq $$

To see if  \stlim\ can be satisfied in realistic experimental
situations, we estimate various quantities using values
typical to crystal growth from solution. As an example,
usual crystal growth from solution has a value of
supersaturation $(c_{\infty}-c_{0})/c_{0}\approx 10^{-2}$.
Further, we use $\gamma\Omega/k_BT=25$\AA, $c_{eq}/c_r
=10^{-2}, x_s=800$\AA, for NaCl crystals at $T=600$K.
This gives an estimate that $d_0\approx 3\times 10^{-4}$,
$\Delta \approx 10^{-4}$, and therefore $d_0/\Delta^2\gg 1$.
Thus \stlim\ is most likely to hold for conventional crystal growth,
implying complete linear stability of straight steps.
However, instability may occur under MBE growth situations
where much higher supersaturation can be achieved, as suggested
in Ref.\ghez.

Although \stlim\ is derived within the
symmetric model, the qualitative conclusion, that there exists
a critical value of line tension above which the step
is linearly stable, is unaltered when asymmetric attachment
kinetics are taken into account.
Only the numerical value of the threshold \stlim\ is modified.
It will be shown later that in a one-sided model with strong
attachement kinetics, the threshold \stlim\ changes to
$d_0\geq \Delta/2$. So step instability is relatively
easier to realize, even for the parameters of conventional
crystal growth quoted above.

\section{Stability of Step Train Under Lateral Fluctuations}

We now examine the linear stability of an infinite train of
equidistant steps. It is useful to distinguish
two types of morphological perturbations:
the lateral fluctuations to the shapes of steps, and
the longitudinal displacement of steps from steady
state positions.
Both kinds of perturbations are accounted for
simultaneously in the stability equation \stab.

In this section we first investigate the linear stability
of steps with respect to fluctuations of the first type,
assuming to this end that the average spacing between
steps is maintained at fixed value $\lambda$.
Still, shape perturbations on different steps need not be
synchronous, but the most unstable situation corresponds to
all steps having in-phase fluctuations. That is,
$\hz^m(k,\omega)=\hz^n(k,\omega)$ for arbitrary $m,n$.
In this case, equation \stab\ yields
 $$\eqalign{
\frac{1}{lM} & \left(\frac{i\omega l^2}{2}+1\right)
\left[\coth \left(M+\frac{1}{l}\right)\frac{\lambda}{2}
     +\coth \left(M-\frac{1}{l}\right)\frac{\lambda}{2}\right] \cr
& +\left[\coth\left(M+\frac{1}{l}\right)\frac{\lambda}{2}
     -\coth\left(M-\frac{1}{l}\right)\frac{\lambda}{2}\right]
     \cr
& =  2\left( \Delta-d_0lk^2\right)
 +\left( \coth \frac{\alpha_+\lambda}{2}
     -\coth \frac{\alpha_-\lambda}{2}\right)
   } \eqn\manystab $$
which can be solved implicitly for the
dispersion relation $\omega=\omega(k)$.
The presence of a zero mode in the spectrum,
due to the translational invariance of the steady state
solution, is evident by explicit inspection: $\omega=k=0$ solves
\manystab. Unfortunately, equation \manystab\
does not afford an explicit analytic solution of $\omega(k)$.
However, one can still demonstrate that
steps become completely linearly stable for strong enough
line tension. The critical value of
the dimensionless capillary length is determined via the
condition
$$\frac{d^2\omega (k)}{dk^2}\bigg|_{k=0}=0  $$
which gives the formula for the threshold line tension:
 $$  d_0\geq
   \frac{\alpha_+-\alpha_-}{(\alpha_++\alpha_-)^2}\left[
  \Delta+
\frac{\alpha_+\lambda e^{\alpha_+\lambda}}{(e^{\alpha_+\lambda}-1)^2}
-\frac{\alpha_-\lambda e^{\alpha_-\lambda}}{(e^{\alpha_-\lambda}-1)^2} \right]
 \eqn\dlimit $$
where $\alpha_{\pm}=\alpha_{\pm}(\Delta,\lambda)$ have been defined
previously and can be solved from the steady state solution
\stead.
Two limits, of large and small step spacing respectively,
are analytically tractable. Specifically, we obtain from \stead\
that
 $$ \alpha_{+}=1+\frac{\Delta}{2}\lambda +..
 \eqn\junkone $$
for $\lambda\ll 1$ and
 $$ \alpha_{+}=
 \sqrt{\frac{1+\Delta}{1-\Delta}}
  \left[1+ \frac{\Delta}{1-\Delta^2}
  \left(e^{-\sqrt{\frac{1-\Delta}{1+\Delta}}\lambda}
       +e^{-\sqrt{\frac{1+\Delta}{1-\Delta}}
\lambda}\right) +...   \right]  \eqn\junktwo  $$
for $\lambda\gg 1$.
Then condition \dlimit\ reduces, in the two limits, to
 $$  d_0\geq \frac{\Delta^2}{360}\lambda^5 + O(\lambda^7)
 \eqno\eq $$
for $\lambda\ll 1$ and
 $$ d_0\geq \frac{\Delta^2\sqrt{1-\Delta^2}}{2}
   - \frac{\Delta\lambda}{2} \left[(1-\Delta)
   e^{-\sqrt{\frac{1-\Delta}{1+\Delta}}\lambda }
  -(1+\Delta)
   e^{-\sqrt{\frac{1+\Delta}{1-\Delta}} \lambda}\right]
             +...   \eqno\eq  $$
for  $\lambda\gg 1$.
Hence the critical value of $d_0$ needed to
linearly stablize straight steps of
finite spacing against lateral perturbations
is always smaller than that required for an isolated step.
This observation is consistent with the intuitive expectation
that smaller $\lambda$ suppresses
lateral fluctuations of the steps, due to
greater overlap of adatom  diffusion fields.
For asymmetric attachement kinetics, this conclusion
remains qualitatively true as well, and has been
reached by previous authors\refmark{\bales}
using quasi-static analyses.

Our analysis is strictly linear, therefore does not address
the inquiry regarding what happens once straight steps
becomes linearly unstable. A natural expectation is that
the steps may adopt profiles corresponding to other
steady state solutions. A recent nonlinear, quasi-stationary
analysis by Misbah and Rappel\Ref\misbah{
C. Misbah and W. Rappel, preprint.} for the
asymmetric model established the existence of
cellular step solutions with a continuum band of possible
wavelengths. Whether or which of these solutions
can be actually selected is intimately related to their
stability, an issue which remains unresolved.

\section{Stability of Step Train Under Longitudinal Fluctuations}

We now investigate the linear stability of an infinite step train
against perturbations in the distances between steps.
In this situation, an instability could lead to the coalescence
or bunching of individual monoatomic steps into
multiple-step bands, a feature frequently observed during
crystal morphological changes.  For simplicity, we assume
that the line tension is large enough so that all steps
remain flat so that we can set $k=0$ in the stability equation.
Arbitrary displacement of steps from their steady state positions
can be decomposed into the normal modes as
$\hz^n(k,\omega)=e^{iq\lambda n}u_q$, with $q$ being the
real wavenumber. Without loss of generality, the wavenumber $q$ can
be further restricted to its first Brillouin zone
$q\in [-\frac{\pi}{\lambda},\frac{\pi}{\lambda}]$.
The stability equation \stab\  then reads
  $$ \sum_{n=-\infty}^{+\infty}
        A_n(i\omega) e^{iq\lambda n}=0   \eqn\bunsym $$
where the coeffients $A_n(i\omega)$ are given by
$$ A_0(i\omega)=\frac{i\omega l^2}{2}+1 -{\hat M}\left(
     \Delta+\frac{1}{e^{\alpha_+\lambda}-1}
 -\frac{1}{e^{\alpha_-\lambda}-1} \right)  \eqno\eq $$
$$ A_{n>0}(i\omega)=\left(\frac{i\omega l^2}{2}+1-
       {\hat M}\right)e^{-({\hat M}
  -1)n\lambda/l} \eqno\eq $$
$$ A_{n<0}(i\omega)=\left(\frac{i\omega l^2}{2}+1+
        {\hat M}\right)e^{({\hat M}
   +1)n\lambda/l} \eqno\eq $$
where ${\hat M}(i\omega)\equiv
lM(k=0,\omega)=\sqrt{l^2+1+i\omega l^2}$.
The frequency $\omega$ is in general complex,
so it is appropriate to separate the dispersion relation into
a real and an imaginary part, $i\omega(q)=\Omega_R(q)+i\Omega_I(q)$.
Positive values of $\Omega_R$ imply exponential amplification
of small perturbations hence linear instability.
Once again one can check by explicit
substitution the presence of a translational zero mode
$\Omega_R(q=0)=\Omega_I(q=0)=0$.

First let us examine in detail the analytically
tractable limit of long-wavelength perturbations
$|q\lambda| \ll 1$, corresponding to dispersion spectrum
near the center of the Brillouin zone.
Taylor expansion of the stability equation gives
in this limit  the relations
$$ \Omega_I(q)=-\frac{\sum nA_n}{\sum A_n'}\, (q\lambda)
  +O(q\lambda)^3  \eqn\bunilim $$
$$ \Omega_R(q)= \frac{\sum A_n''}{2\sum A_n'}\Omega^2_I
   +\frac{\sum nA_n'}{\sum A_n'}(q\lambda)\Omega_I
   +\frac{\sum n^2A_n}{2\sum A_n'}(q\lambda)^2 +O(q\lambda)^4
   \eqno\eq $$
where primes denote derivatives with respect to $i\omega$
evaluated at $i\omega=0$.
Note that in the long wavelength limit, $\Omega_I(q)$
is linear but $\Omega_R(q)$ is quadratic in the wavenumber.
Formula \bunilim\ can be evaluated to give
$$ \Omega_I(q)= \frac{(\alpha_++\alpha_-)
      (\alpha_+\sinh^{-2}\frac{\alpha_+\lambda}{2}+
      \alpha_-\sinh^{-2}\frac{\alpha_-\lambda}{2})
}{4\Delta l^2-\lambda
 \alpha_+\sinh^{-2}\frac{\alpha_+\lambda}{2}+\lambda
      \alpha_-\sinh^{-2}\frac{\alpha_-\lambda}{2}
}\,q\lambda . \eqno\eq $$
The expression for the real mode is much more
complicated and is not displayed here.
Despite the complexity, behavior in the limits of large
and small step spacings can still be extracted, using the
expansions \junkone\ and \junktwo. The results are, for
$\lambda\gg 1$,
$$ \Omega_I(q)=\frac{2\Delta}{(1-\Delta^2)(1+\Delta)}\left(
    e^{-\sqrt{\frac{1-\Delta}{1+\Delta}}
\lambda}+ \frac{1+\Delta}{1-\Delta}e^{
-\sqrt{\frac{1+\Delta}{1-\Delta}}\lambda }+...\right)
   q\lambda    \eqno\eq $$
$$ \Omega_R(q)=\frac{-\Delta^2}{(1-\Delta^2)(1+\Delta)} \left(
    e^{-\sqrt{\frac{1-\Delta}{1+\Delta}}
\lambda}- \frac{1+\Delta}{1-\Delta}e^{
-\sqrt{\frac{1+\Delta}{1-\Delta}}\lambda }+...\right)
 (q\lambda)^2   \eqno\eq $$
and for $\lambda\ll 1$,
$$ \Omega_I(q)=[
    \Delta+O(\lambda^2) ]q\lambda \eqno\eq $$
$$ \Omega_R(q)=\left[-\frac{\Delta^2(1+3\Delta^2)}{12}
    \lambda^2 +O(\lambda^4)\right](q\lambda)^2 . \eqno\eq $$
These asymptotic expansions are obtained from the full linear
stability equation, hence involve no approximations.
We note in particular that our result for $\Omega_I(q)$
agrees with that of Bennema and
Gilmer\rlap,\Ref\bennema{P. Bennema
and G.H. Gilmer, in {\sl Crystal Growth: An
Introduction}, edited by P. Hartman (North Holland, Amsterdam,
1973).} obtained in a simple model assuming a step velocity
depending upon only nearest terrace widths. Most importantly,
however, in contrary to a pure neutral mode
$\Omega_R(q)=0$ found in their approximate model, we
obtain a weakly stable mode $\Omega_R\leq 0$.

While the magnitude of $\Omega_R(q)$ reflects the typical decay or
amplification rate of perturbations, $\Omega_I(q)$ is related to
the wave velocity for the propagation of disturbances.
When the wavelength of disturbances is large compared with
the step spacing, i.e., $q\lambda\ll 1$, the group velocity
of disturbances coincides with the phase velocity since
$\Omega_I(q)$ is linear in $q$.  Both are given
by $v=-\Omega_I/q$ in the co-moving frame.
We immediately see that for small
step spacing $\lambda\ll 1$, $v=-V$, so the disturbances remain
stationary in the lab frame. On the other hand, for large step
separation $\lambda\gg 1$, $v=0$, and the disturbances
propagate along with the steps.

The exact linear stability spectrum \bunsym\
over the complete Brillouin zone can only be solved numerically.
However, we can attempt an approximate solution utilizing a
quasi-static approximation. As is usually done,
this approximation assumes that the diffusion field adjusts quickly
in response to the displacement of the steps.
It amounts to neglecting the frequency dependence, or the memory effect,
in the Green's function: ${\hat M}(i\omega)=\sqrt{l^2+1+i\omega l^2}
 \approx \sqrt{l^2+1}$.
Now \bunsym\ can be easily solved over the whole Brillouin zone to
yield
$$\eqalign{
\Omega_R(q)= \frac{2}{l^2} & \left[
   \left(
  \Delta+\frac{1}{e^{\alpha_+\lambda}-1}   \right . \right. \cr
& \left . \left .  -\frac{1}{e^{ \alpha_-\lambda}-1} \right)
  \sqrt{1+l^2}\coth \lambda \sqrt{1+ \frac{1}{l^2}}
   -1  \right](1-\cos q\lambda)} , \eqn\symquar$$
$$ \Omega_I(q)=
\frac{2\sqrt{1+l^2}\sinh\lambda/l}
  {l^2\sinh \lambda\sqrt{1+1/l^2}}
\left[ \coth\frac{\lambda}{l}
 -\Delta -\frac{1}{e^{\alpha_+\lambda}-1}+\frac{1}{e^{
   \alpha_-\lambda}-1}
 \right] \sin q\lambda   .   \eqn\symquai $$
The quasi-static dispersion relation is shown, in dashed lines,
along with the numerically determined exact spectrum, in Fig.3.
The most important feature is that $\Omega_R(q) \leq 0$ for
all $q$, suggesting complete linear stability of the
equidistant step train against fluctuations in step spacings.
\FIG\figthree{Dispersion relation of an infinite train
of parallel steps against perturbations in step spacings,
in the symmetric model. Solid lines are obtained
numerically from \bunsym, and dashed lines
correspond to the quasi-static spectrum \symquar\ and
\symquai. Curves in the upper figure, from top down,
are for $\lambda=0.5, 1$, and $2$. Order of curves is reversed
in the lower part.}

To suppress the frequency dependence in
$\hm(i\omega)$ when adopting the quasi-static approximation, we need
$|\omega|\ll 1$. From the quasi-static spectrum, it can be
seen that typically $\Omega_I=O(\Delta)$ and $\Omega_R=O(\Delta^2)$.
Hence the condition of small supersaturation
$\Delta \ll 1$ is necessary.
Another, more subtle, source of inaccuracy may occur in the
quasi-static procedure, if the magnitudes of $\Omega_R$
and $\Omega_I$ have large disparity between them.
This seems to explain the discrepancies between the quasi-static
and the exact spectra near the center of the Brillouin zone
for small $\lambda$ in Fig.3.

\chapter{GROWTH AND EVAPORATION WITH ASYMMETRIC ATTACHMENT KINETICS}

In the preceding discussions, steps are assumed to be ideal
line sources/sinks near which a continuous adatom density is
maintained at the local equilibrium value.
This picture must be modified, in several respects, to
accommodate to more realistic growth situations.
First of all, effects such as the finite rate of adatom exchange
at steps, insufficient concentration and non-ideal distribution
of sink sites on steps, all lead to deviations of adatom density
at the steps from the equilibrium value.
In other words, growth rate is no longer
solely controlled by adatom transport, but also by
interface kinetics.  Secondly,
atomic exchanges on different sides of steps
involve different kinetic barriers and are therefore nonequivalent.
This asymmetry in attachment kinetics was pointed out by
Schwoebel and Shipsey\rlap,\refmark{\schwoebel}
and discussed extensively in subsequent literatures. Lastly,
interface kinetics usually exhibit pronounced anisotropy.

The effect of interface kinetics is usually taken into account
by assuming an elevation of $c_{step}$ from
the equilibrium value linearly proportional to
the adatom flux.
Accordingly boundary condition \bound\ is replaced by
 $$ c_{step\pm}-c_{0}\left(1+\frac{\gamma\Omega}{k_BT}
\kappa\right) =\beta_{\pm} D (\nabla c) \cdot {\bm n}_{\pm} \eqn\modbound $$
where differences in the coefficients $\beta_{\pm}$ measure the
asymmetry of attachment kinetics.
We have for simplicity neglected effect of anisotropy.
In this section, we consider a situation with the most asymmetry,
assuming local equilibrium near lower terraces but complete
inhibition of adatom exchange from upper terraces.
This amounts to setting $\beta_+=0$, $\beta_-=\infty$
in \modbound.
We further distinguish between two particular cases of
crystal growth and evaporation, respectively, shown schematically
in Fig.4.
\FIG\figfour{Schematic representation of step flow during:
(a) crystal growth and, (b) evaporation with asymmetric
attachement kinetics.}
Again working in the moving frame with velocity $V$ in
the $z$-direction, the boundary conditions \eimtwo\ and
\eimthree\ are modified to the following dimensionless form
 $$\left(V+\frac{d\zeta^n}{dt}\right)dx = -(\nabla u)^n_+
 \cdot d\vecsigma^n;
\qquad  (\nabla u)_-\cdot d\vecsigma^n=0  \eqno\eq $$
 $$ u_+(x,\zeta^n,t)= \Delta -d_0 \kappa\{\zeta^n\}  \eqno\eq $$
for the case of growth  and
 $$\left(V+\frac{d\zeta^n}{dt}\right)dx = (\nabla u)^n_-
    \cdot d\vecsigma^n;  \qquad (\nabla u)_+\cdot d\vecsigma^n=0   \eqno\eq $$
 $$ u_-(x,\zeta^n,t)= \Delta -d_0 \kappa\{\zeta^n\}  \eqno\eq $$
for the case of evaporation. The bulk diffusion equation
\movfr\ remains intact in both cases.
Note however that for evaporation, the quantity $\Delta$ $(>0)$
is identified with the dimensionless undersaturation and
the dimensionless diffusion field in this case is defined according
to $u=(c-c_{\infty})/c_r$ in contrast with \defofu.
The above models are analogies of the one-sided models in
solidification.

\section{The Case of Growth}
Integro-differential equations for the step displacement are
derived using the same method as before by applying
Green's theorem to the domain shown in Fig.4.
Since adatom density will not be continuous across a step,
twice as many equations are needed compared to the symmetric case.
A static version of the Green's function method was recently
used by Misbah and Rappel\refmark{\misbah} in a bifurcation
study of the steady state solution of cellular steps.

Consider an arbitrary point $p\in \Lambda_1$
on the terrace between steps A and B in Fig.4. Green's theorem
gives
 $$\eqalign{
   u(p) & =\int_A dx_1G(p|p_1)\left[
     V(1-u^A_+(p_1))+\frac{d\zeta^A_1}{dt_1}\right]
   +\int_A d\vecsigma_1 u^A_+(p_1)\cdot
    \nabla_1 G(p|p_1)   \cr
 & + \int_B d\vecsigma_1 u^B_-(p_1)\cdot
    \nabla_1 G(p|p_1) +V\int_B dx_1u^B_-(p_1) G(p|p_1) } \eqno\eq $$
where to save notation the integration over $dt_1$ is implied
in the above expression.
When we let the point $p$ approach step A, the second term on
the r.h.s. with $\nabla_1 G$ develops an integrable singularity.
After separating the contribution of this singularity, we obtain
$$ \eqalign{
u^A_+ &(\zeta^A,t)= 2\int_B d\vecsigma_1
u^B_-(p_1)\cdot \nabla_1 G(p|p_1) +2V
    \int_B dx_1u^B_-(p_1) G(p|p_1) +   \cr
 & \int_A dx_1\left[
V(2-u^A_+(p_1))+2\frac{d\zeta^A_1}{dt_1}
  +(\zeta^A-\zeta^A_1-\frac{d\zeta^A_1}{dx_1}(x-x_1))
  \frac{u^A_+(p_1)}{t-t_1}\right] \cr
 & \times G(p|p_1) .  } \eqn\groa  $$
Another equation is similarly obtained letting $p$ approach step B,
 $$ \eqalign{
u^B_- &(\zeta^B,t)=2\int_A d\vecsigma_1
u^A_+(p_1)\cdot \nabla_1 G(p|p_1)
   +2 \int_A dx_1 \left[V(1-u^A_+)+\frac{d\zeta^A_1}{dt_1}\right]
 G(p|p_1) \cr  & + \int_B dx_1G(p|p_1)\left[
V-(\zeta^B-\zeta^B_1-\frac{d\zeta^B_1}{dx_1}(x-x_1))
  \frac{u^B_-(p_1)}{t-t_1}\right] .  } \eqn\grob  $$

Once again, linearization of \groa\ and \grob\ leads
to the stability equation. Here we first briefly mention
the linear stability of an isolated straight step
under lateral shape perturbations. In our formalism,
it turns out the dispersion relation can again be
solved closely, from the relation,
 $$ i\omega l^2+2=(\Delta-d_0lk^2)(lM+1)+
    \Delta l(k^2l+\sqrt{1+l^2}M-lM^2) \eqno\eq $$
where $M(k,\omega)=\sqrt{k^2+1+1/l^{2}+i\omega}$.
The actual expression for $\omega(k)$ is lengthy and not shown
here. We only point out that the threshold line tension
needed to linearly stabilize the
straight step is now given by
 $$  d_0 \geq \frac{\Delta\sqrt{1-\Delta}}{2-\Delta} , \eqno\eq $$
while for smaller $d_0$ the step is unstable to long-wavelength
perturbations.  Thus suppressing adatom attachment from
the upper terrace makes the step
more susceptible to shape fluctuations.  Further, it can be shown
that the effect of finite step spacing $\lambda$ is stablizing, which
merely reduces the above threshold to smaller
values. The calculation is straightforward and
reproduces the results of previous authors\rlap.\refmark{\bales}

Below, our major attention will be paid to the longitudinal
modes of an infinite step train.
For straight steps, the equations of motion are much simpler.
Since $u^A_+=\Delta$ and $\zeta(x,t)$ is
independent of $x$, the integration over $dx_1$ can be
carried out explicitly. And the equations for step A and B are:
 $$\eqalign{
   \Delta= &\int dt_1
\left[V(2-\Delta)+  2\frac{d\zeta^A_1}{dt_1}\right]
 G(\zeta^A,t|\zeta^A_1,t_1) \cr
 &  +2\int dt_1 u^B_-(t_1)
  [VG(\zeta^A,t|\zeta^B_1,t_1)-\nabla_{1z}G(\zeta^A,t|\zeta^B_1,t_1)
  ]  }  \eqn\flata $$
 $$\eqalign{
   u^B_- & (t)= \int dt_1
\left[2V(1-\Delta)+2\frac{d\zeta^A_1}{dt_1}\right]
G(\zeta^B,t|\zeta^A_1,t_1) \cr &  +2\Delta\int
dt_1\nabla_{1z}G(\zeta^B,t|\zeta^A_1,t_1)+V\int
dt_1u^B_-(t_1)G(\zeta^B,t|\zeta^B_1,t_1) } \eqn\flatb $$
with the reduced Green's function
 $$ G(zt|z_1t_1)=
\frac{e^{-(t-t_1)}}{\sqrt{4\pi (t-t_1)}}
\exp \left[-\frac{(z-z_1+V(t-t_1))^2}{4(t-t_1)} \right]. \eqno\eq $$
Uniformly moving steady states are zeroth order solutions of
\flata\ and \flatb. Setting $\zeta^A=0, \zeta^B=\lambda$,
we obtain
 $$ u_-= e^{-\alpha_+\lambda}(\alpha_+^2+\Delta-1)
 = e^{\alpha_-\lambda} (\alpha_-^2 +\Delta-1),   \eqno\eq $$
from which both the velocity of the step and the adatom density
at the upper side of the steps can be calculated.
And the magnitude of discontinuity in adatom density across the step
equals $\Delta-u_-$.

The linear stability equation, derived from \flata\ and \flatb\
by linearization, follows the same structure as \bunsym, but with only
two terms $n=0, 1$ in the summation. It reads
$$ \eqalign{
& \frac{\hm-1}{\hm+1}e^{(\hm-1)\lambda/l}
     \left[i\omega l^2  +2-\Delta -2\hm+\Delta\hm
  \sqrt{1+l^2}\right] +e^{-(\hm+1)\lambda/l}\times \cr
& \left[ i\omega l^2+2- \Delta +2\hm+\Delta\hm^2
   \right]
 =u_-e^{iq\lambda}\hm (\hm +\alpha_-l-1) }  \eqn\bungrow $$
where as before ${\hat M}(i\omega)=\sqrt{1+l^2+i\omega l^2}$.
It is easier in this case to extract the dispersion relation near
the center of the Brillouin zone.
We find, to leading order,  that
$$ \Omega_I(q)=\frac{2\Delta(2-\Delta)^3}{(1-\Delta)^2(4-2\Delta-\Delta^2)}
 e^{-\frac{2-\Delta}{\sqrt{1-\Delta}}\lambda }
    q\lambda  \eqno\eq $$
$$\Omega_R(q) =
\frac{-\Delta(2-\Delta)^3}{(1-\Delta)^2(4-2\Delta-\Delta^2)}
 e^{-\frac{2-\Delta}{\sqrt{1-\Delta}}\lambda } (q\lambda)^2
  \eqno\eq $$
in the limit of large step separation $\lambda \gg 1$ and
$$ \Omega_I(q)=\left[\Delta +O(\lambda^2)\right]q\lambda \eqno\eq $$
$$ \Omega_R(q)
=-\left[\frac{\Delta}{2}+\Delta^2+
  \frac{3\Delta^3}{32}+ O(\lambda^2)\right](q\lambda)^2
   \eqno\eq $$
in the limit of narrow step trains.
The spectrum solved under the quasi-static approximation
also takes a particularly simple form:
$$ \Omega_I(q)= \frac{
    u_-^2e^{2\lambda/l}}{\Delta}
    \sin q\lambda  \eqno\eq $$
$$ \Omega_R(q)=-\frac{u_-^2e^{2\lambda/l}}{\Delta}
(1-\cos q\lambda) .    \eqno\eq $$
Note that both $\Omega_R$ and $\Omega_I$ are of the the same
order $O(\Delta)$.
In Fig.5, a comparison of the exact spectrum with the
quasi-static one suggests the latter to be a very good
approximation.  Since $\Omega_R(q)\leq 0$ for all $q$,
the infinite step strain is linearly stable against
perturbations in step spacings.
\FIG\five{Dispersion relation of an infinite train of parallel
steps against perturbations in step spacings during crystal
growth, in the one-sided model. Solid lines represent
the numerically solved exact spectrum, and dashed lines
correspond to the quasi-static spectrum.
Curves, in increasing amplitude, are for $\lambda=2, 1$, and $0.5$. }

\section{The Case of Evaporation}

With little modification, procedures used in the last subsection
can be applied to the case of crystal evaporation.
We will not repeat the analyses, but only quote the results.
For an isolated straight step, the linear stability spectrum for
lateral fluctuations is the closed solution of
 $$ i\omega l^2+2=(\Delta-d_0lk^2)(lM-1)-
    \Delta l (k^2l+\sqrt{1+l^2}M-lM^2) .  \eqno\eq $$
It follows that $i\omega$ is pure real and $i\omega (k)\leq 0$,
for arbitrary $d_0$ and $k$.  So a straight step
is completely linear stable against shape fluctuations. This
conclusion remains valid for equidistant step trains with finite
$\lambda$ as well.

The equations of motion for parallel steps resemble those
in the case of growth and read
$$\eqalign{ \Delta= &\int dt_1
\left[V(2+\Delta)+ 2\frac{d\zeta^B_1}{dt_1}\right]G(\zeta^B,t|\zeta^B_1,t_1)
  \cr
 &  -2\int dt_1 u^A_+(t_1)
[VG(\zeta^B,t|\zeta^A_1,t_1)-\nabla_{1z}G(\zeta^B,t|\zeta^A_1,t_1)
  ] ,   }  \eqno\eq $$
$$\eqalign{ u^A_+ & (t)= \int dt_1
\left[2V(1+\Delta)+2\frac{d\zeta^B_1}{dt_1}\right]
 G(\zeta^A,t|\zeta^B_1,t_1)
  \cr &  -2\Delta\int
dt_1\nabla_{1z}G(\zeta^A,t|\zeta^B_1,t_1)-V\int
dt_1u^A_+(t_1)G(\zeta^A,t|\zeta^A_1,t_1) .  } \eqno\eq $$
Stationary state adatom density and step velocity can be
solved from
 $$ u_+= e^{-\alpha_-\lambda}(1+\Delta-\alpha_-^2)
 = e^{\alpha_+\lambda} (1+\Delta-\alpha_+^2) \eqno\eq $$
and the linear stability equation is
 $$ \eqalign{
& \frac{\hm+1}{\hm-1} e^{(\hm+1)\lambda/l}
  \left[i\omega l^2  +2+\Delta +2\hm-\Delta\hm
 \sqrt{1+l^2}\right] +e^{-(\hm-1)\lambda/l}\times \cr
& \left[ i\omega l^2+2+\Delta -2\hm-\Delta\hm^2
   \right] =-u_+ e^{-iq\lambda}\hm (\hm +\alpha_+l+1) .  }\eqno\eq $$
The quasi-static dispersion relation,
$$\Omega_I(q)= \frac{u_+^2e^{-2\lambda/l}}{\Delta}
    \sin q\lambda  \eqno\eq $$
$$ \Omega_R(q)= \frac{u_+^2e^{-2\lambda/l}}{\Delta}
    (1-\cos q\lambda), \eqno\eq $$
is displayed in Fig.6 with the exact spectrum.
We immediately observe that $\Omega_R(q) \geq 0$ for
all values of $q$. Therefore, while the infinite step train
is linearly stable against lateral shape fluctuations, it is
unstable to perturbations in step spacings, of all wavelengths.
\FIG\six{Dispersion relation of an infinite train of parallel
steps against perturbations in step spacings during
evaporation, in the one-sided model. Numerically solved
exact spectrum is shown in solid lines, and the quasi-static
spectrum is shown in dashed lines. Curves, in increasing amplitude,
correspond to $\lambda=2, 1$, and $0.5$. }

\chapter{MORPHOLOGICAL FLUCTUATIONS OF STEPS}

At finite temperatures, a monoatomic step on a singular
crystal surface contains kink excitations.
As the temperature is raised, the density of kink
and other short-range structural excitations increases
and the step can become molecularly rough.
In fact, in our previous considerations of step kinetics,
molecular roughness of steps is necessary
for them to behave as good sources/sinks of adatoms
so that the continuum description applies.
Moreover, roughness of steps can persist to macroscopic
scales comparable to step extension,
since a step is always {\it thermodynamically} rough
at finite temperature meaning that the amplitude of
fluctuations diverges with the system size.
Beside contribution from equilibrium fluctuations,
step roughness is also affected by nonequilibrium
properties of the growth process itself, for example, by the
amplification of noises by the Mullins-Sekerka instability as
we shall show below.

Previous studies of step morphology were mostly
confined to treatments of the equilibrium fluctuations on
weakly interacting steps\rlap.\Ref\williams{Equilibrium
properties of steps are recently reviewed in E.D. Williams
and N.C. Bartelt, \journal Science &251&393(91).}
In addition, the effect of thermal and other
excitations on the terraces
are routinely neglected.  It is not until recently,
that attention has begun to be paid to step fluctuations under
non-equilibrium situations. In one such study
by Uwaha and Saito\rlap,\Ref\uwaha{M. Uwaha and Y. Saito,
\journal Phys. Rev. Lett.
&68&224(92); \journal Surf. Sci. &283&366(93).}
step roughness is examined by adding to the linear
stability equation a hypothetical stochastic noise
term representing the cumulative effect of all thermal
fluctuations. The origin of their choice of the randomness,
however, remains elusive.
Another study, by Salditt and Spohn\Ref\spohn{
T. Salditt and H. Spohn, \journal Phys. Rev. E. &47&3524(93).}
on the time-dependent step roughness, took into account bulk random
noise from the terraces but neglected noise
contributions from excitations on the
steps. Neither investigation considers multiple steps.

In this section, non-equilibrium step roughness
is investigated using a Langevin formalism.
Our approach is an extension of the general theoretical framework
for the study of morphological fluctuations of
solidification fronts at or near equilibrium, developed
by Langer\Ref\jimfluc{J.S. Langer, \journal Phys. Rev. A
&36&3350(87); J.A. Warren and J.S. Langer,
\journal Phys. Rev. E &47&2702(93).}
and Karma\rlap,\Ref\karma{A. Karma, \journal Phys. Rev. Lett.
&70&3439(93); and unpublished.}
based on the procedures introduced by
Cherepanova\rlap.\Ref\cherepanova{T.A. Cherepanova,
\journal Dokl. Akad. Nauk SSSR &226&1066(76) [\journal
Sov. Phys. Dokl. &21&109(76)].}
In this formalism, non-equilibrium, hydrodynamical fluctuations
of the phase boundaries are evaluated as responses to the
Langevin forces representing local thermodynamical fluctuations.
We demonstrate that this Langevin formalism provides a consistent
description of the kinetic roughness of steps. The effect of
finite step spacing on step roughness, and the correlation
between fluctuations on different steps, can all be conveniently
evaluated.  Our approach is limited in several aspects.
The first and foremost limitation is related to the
unresolved fundamental question of statistical mechanics
of whether the Langevin formalism itself is applicable in
treating statistical fluctuations of nonequilibrium systems.
Second, our calculation is carried out in the small amplitude limit and
allows neither overhangs nor crossings of steps.
Similarly, no interactions between steps
beside the pure diffusive coupling is taken into account.

In the Langevin approach, stochastic noises are introduced to
the diffusion equation and to the Gibbs-Thompson boundary
condition as
 $$ \frac{\pa u\rt}{\pa t} =\nabla^2 u -u +q\rt \eqno\eq $$
 $$ u(x,\zeta^n,t)=\Delta-d_0\kappa\{\zeta^n\}-\beta
     v_n\{\zeta^n\}+\eta\{\zeta^n\}  \eqno\eq $$
where the two independent Langevin forces $q$ and $\eta$
mimic fluctuations in adatom density on the terraces and
near the steps respectively.
The noises are assumed to be Gaussian distributed with zero mean
and variances
  $$ \VEV{q\rt q(\vecr',t')}=2\Gamma (1-\nabla^2)\delta({\bm r}
     -{\bm r}')\delta(t-t')  \eqn\qvar $$
  $$ \VEV{ \eta (x,t)\eta(x',t')}=2\beta\Gamma
   \delta(x-x')\delta(t-t') .  \eqn\qvartwo $$
Note that the two-dimensionally distributed noise $q$ comprises
two contributions: one coming from the randomness in the impinging flux
of atoms from the vapor, and the other from the adatom terrace
diffusion processes itself.  The magnitude $\Gamma$, as yet
unspecified, will be fixed by the requirement that the method
recovers correctly the equilibrium fluctuations.

Restricting ourselves to an infinite equidistant step train
in the symmetric model, we write the stochastic equation of motion
for the $n$th step as
$$\eqalign{
 \Delta- d_0\kappa-& \beta v_n
     +\eta\{\zeta^n\}-\sigma(x,\zeta^n,t)   \cr
 &=\sum_{m} \int dt_1 dx_1 G\left(x, \zeta^n,t|x_1,\zeta^m_1,t_1
   \right)\left(V+ \frac{d\zeta^m_1}{dt_1}\right) }
    \eqno\eq $$
where the two-dimensional noise is projected as
  $$ \sigma(x,\zeta^n,t)=\int dt_1d\vecr_1
      G(x,\zeta^n,t|\vecr_1,t_1)q(\vecr_1,t_1) .  \eqno\eq $$
Considering small amplitude fluctuations around straight
steps we linearize the above equation in the Fourier
representation. The step profiles in response to the noises
satisfy
 $$ \sum_{m=-\infty}^{+\infty} A_{n-m}(k,\omega) \hz^m(k,\omega)=
     \sigma^n(k,\omega)-\eta^n(k,\omega)   \eqn\toep $$
where coefficients $A_{n-m}$ only depend on the difference $n-m$:
 $$ A_0=\frac{1}{l}\left(\Delta+
\frac{1}{e^{\alpha_+\lambda}-1}-\frac{1}{e^{\alpha_-\lambda}-1}
 -\frac{1}{Ml}\right) -i\omega \left( \frac{1}{2M}+\beta\right)
  -d_0k^2     \eqno\eq $$
 $$ A_{n-m}=-\left(\frac{i\omega }{2M}+\frac{1}{Ml^2}+\frac{1}{l}
   \right) e^{-(M+1/l)(n-m)\lambda},  \qquad   n>m     \eqno\eq $$
 $$ A_{n-m}=-\left(\frac{i\omega }{2M}+\frac{1}{Ml^2}-\frac{1}{l}
   \right) e^{(M-1/l)(n-m)\lambda},  \qquad  n<m   \eqno\eq $$
with $M(k,\omega)$ as previously defined. Also, we have
 $$ \sigma^n(k,\omega)= \int_{-\infty}^{+\infty}
        dz_1 e^{-(z_1-n\lambda)/l-|z_1-n\lambda|M}
    \frac{q(k,\omega,z_1)}{2M(k,\omega)}.  \eqno\eq $$
The infinite set of equations \toep\ have the Toeplitz form
and can be solved straightforwardly using Fourier series. Defining
 $$ {\Cal A} (\theta) = \sum_{n=-\infty}^{\infty}
        A_n e^{in\theta},   \, \, \,
{\Cal Z} (\theta)=\sum_{n=-\infty}^{\infty}
       \hz^n e^{in\theta},   \eqno\eq $$
 $$ {\Cal S}  (\theta) = \sum_{n=-\infty}^{\infty}
        (\sigma^n -\eta^n) e^{in\theta} ,  \eqno\eq $$
we have
 $$ {\Cal A}(\theta){\Cal Z}(\theta)=
    {\Cal S}(\theta).  \eqno\eq $$
Inverting the Fourier series, we obtain
 $$ \hz^n(k,\omega) = \int_{-\pi}^{\pi}
        \frac{d\theta}{2\pi} \frac{{\Cal S}(\theta)
e^{-in\theta}}{{\Cal A} (\theta)} , \eqno\eq $$
which leads, noticing that
$\VEV{\sigma^n\sigma^m}$ only depends on $n-m$ and that
$\VEV{\eta^n\eta^m}=\VEV{\eta^0\eta^0}\delta_{nm}$,
to the result
 $$ \VEV { \hz^m(k\omega)\hz^0(k'\omega')}
  = \int_{-\pi}^{\pi}
        \frac{d\theta}{2\pi} \frac{
  \VEV{\eta^0(k\omega)\eta^0(k'\omega')}+\sum_{n} e^{in\theta}
  \VEV{\sigma^n(k\omega)
    \sigma^0(k'\omega')}
   }{ {\Cal A}_{k\omega}(\theta)
     {\Cal A}_{k'\omega'}(-\theta) } e^{-im\theta} . \eqn\flux $$
This expression is the fluctuation spectrum for
the infinite step train in response to stochastic
noises from both the terraces and the steps.
The stationary fluctuation spectrum can be evaluated
from \flux\ by integrating out the arguments
$k',\omega'$ and $\omega$:
 $$ \VEV{\hz^m_k\hz^0_{-k}}=\int \frac{d\omega d\omega' dk'}{
   (2\pi)^3} \VEV { \hz^m(k\omega)\hz^0(k'\omega')} . \eqno\eq $$

To proceed further with \flux, we shall adopt the quasi-static
approximation as before by neglecting the $\omega$ dependence in
$M(k,\omega)$. Consequently, the noise
correlators can be evaluated:
 $$\VEV{\sigma^n(k\omega)\sigma^0(k'\omega')}
   =\Gamma (2\pi)^2\delta(k+k')\delta(\omega+\omega')
  e^{-|n|\lambda M}\frac{1+\sinh n\lambda/l}{M} , \eqno\eq $$
$$\VEV{\eta^0(k\omega)\eta^0(k'\omega')}
   =2\beta\Gamma (2\pi)^2\delta(k+k')\delta(\omega+\omega')
.   \eqno\eq $$
After some algebra, we find the general expression for
the stationary fluctuation spectrum of steps
$$ \VEV{\hz^m_k\hz^0_{-k}}= \frac{\Gamma}{\pi}
 \int_{-\pi}^{\pi} \frac{d\theta e^{-im\theta} a(\theta)}
{|b(\theta)c(-\theta)+b(-\theta)c(\theta)|}   \eqn\stge $$
where
$$ a(\theta)=2\beta M+ \frac{\sinh M\lambda}{\cosh M\lambda -\cos
\theta}  \eqno\eq $$
$$\eqalign{
 b(\theta)=& \frac{1}{l}\left(
\Delta-d_0lk^2+\frac{1}{e^{\alpha_+\lambda}-1}-
              \frac{1}{e^{\alpha_-\lambda}-1}-\frac{1}{Ml}
 \right) \cr
 & -\frac{1}{Ml^2}\left( \frac{1+Ml}{e^{(M+1/l)\lambda-i\theta}-1}
  +\frac{1-Ml}{e^{(M-1/l)\lambda+i\theta}-1} \right) } \eqno\eq $$
$$c(\theta)=
1+2\beta M + \frac{1}{e^{(M+1/l)\lambda-i\theta}-1}+\frac{1}{e^{
(M-1/l)\lambda+i\theta}-1}       \eqno\eq $$
and the quasi-static $M=\sqrt{1+k^2+1/l^2}$.
Equation \stge\ is the main result of this section.

First it is important to examine what \stge\ suggests of
the equilibrium fluctuations.
Setting $\Delta=0, l=\infty$,
we have $a(\theta)=c(\theta)$, $b(\theta)=-d_0k^2$ and hence
the  equilibrium fluctuation spectrum
 $$ \VEV{\hz^m_k\hz^0_{-k}}_{eq}=\frac{\Gamma
\delta_{m0}}{d_0k^2} .  \eqno\eq $$
We observe that the step roughness in equilibrium is
independent of the step spacing $\lambda$.
And apparently, in equilibrium, fluctuations on different steps
also decouple since $\VEV{\hz^m_k\hz^0_{-k}}_{eq}=0$ for
$m\neq 0$.

Under non-equilibrium growth situations, however,
fluctuations on different steps become correlated
since in general $\VEV{\hz^m_k\hz^0_{-k}}\neq 0$ for $m\neq 0$.
Furthermore, equation \stge\ exhibits the familiar
divergence of fluctuations at the onset of Mullins-Sekerka
instability when $b(\theta=0)=0$.
To see this more clearly, we consider the simple limit of
a single step corresponding to $\lambda=\infty$.
In this case the kinetic roughness spectrum of the step
reduces to
$$\VEV{\hz^0_k\hz^0_{-k}}=
   \frac{\Gamma}{|d_0k^2 -\Delta/l+1/Ml^2|} \eqno\eq $$
where the denominator is proportional to the quasi-static
dispersion relation of an isolated step.
Extensions of the present calculation to include
energetic interactions between steps and to incorporate
asymmetric attachment kinetics may be warranted for
direct comparison with experimental observations.

\chapter{COLLECTIVE MOTION OF STEPS --- THE PHASE
FIELD METHOD}

A rich variety of collective behavior of steps are
commonly encountered during changes of crystal surface
topography. Steps are seldom uniformly spaced;
pairwise grouping of
steps, bunching of steps into macro-steps with
height of multiple atomic units, collision and
annihilation of step with antisteps, have all been
observed in various surfaces of, for example,
alkali halides\rlap,\Ref\keller{K.W. Keller,
\journal Metall. Trans. &22A&1299(91); H. Bethge \etal,
\journal J. Crys. Growth &48&9(80).}
semiconductors\rlap,\Ref\latyshev{
A.V. Latyshev, et al, \journal Surf. Sci. &213&157(89);
A.V. Latyshev, A.B. Krasilnikov and A.L. Aseev,
\journal Appl. Surf. Sci. &60&397(92); E. Bauser and
H. Struck, \journal Thin Solid Films, &93&185(82).}
and metals\rlap.\Ref\hsu{T. Hsu, \journal
Ultramicroscopy, &11&167(83);
S. Ogava, et al, \journal J. Vac. Sci. Tech. &5&1735(87).}
To address these effects, fully time-dependent treatments
of the collective movement of steps are necessary.

Theoretical analysis of the time-dependent
multiple step kinetics turns out to be formidable.
The most well known continuum approach is the kinematic
wave theory of Frank\Ref\kinematic{
F.C. Frank, in {\sl Growth and Perfection of Crystals},
eds. R.H. Doremus, B.W. Roberts and D. Turnbull, Wiley,
New York, 1958; N. Cabrera and D.A. Vermilyea,
{\sl ibid}; A.A. Chernov, \journal Sov. Phys. Uspekhi
&4&116(61).}
using the method of characteristics\rlap.\Ref\taylor{
On the application of the method of characteristics to
crystal growth, see J.E. Taylor, J.W. Cahn and C.A.
Handwerker, \journal Acta Metall. Mater, &40&1443(92).}
Frank's theory treats bundles of steps as basic entities and
monitors the temporal changes of the average step density.
It does not trace the position of individual steps and
is therefore more coarse-grained than the continuum BCF model.
However, the method is found particularly useful in describing
large scale changes of surface morphology involving
macro-steps. Time evolution of step spacings
in finite trains of steps has been discussed by Mullins and
Hirth\rlap,\Ref\mullinshirth{
W.W. Mullins and J.P. Hirth, \journal J. Phys. Chem.
Solids, &24&1391(63).}
based on the assumption that the step velocity
is a function of only two adjacent step intervals.
Both approaches apply to straight steps only and do not
allow lateral variation of step profiles.
Other time-dependent studies\Ref\weeks{For example,
D. Kandel and J.D. Weeks, \journal
Phys. Rev. Lett. &69&3758(92);
R. van Rosmalen and P. Bennema, \journal J. Crys. Growth
&32&293(76); C. van Leeuwen,
R. van Rosmalen and P. Bennema, \journal
Surf. Sci. &44&213(74).} of step motion either follow
the spirit of  Mullins and Hirth, or
resort to explicit microscopic Monte Carlo simulations of the
growth processes.

Meanwhile, the attempt towards a direct numerical
solution of the original Stefan problem \dimone\ to \dimthree\
is also difficult, hindered by the need to
track explicitly all moving boundaries.
This difficulty was highlighted in studies of
dendritic solidification\rlap,\refmark{\nigel}
and can be
partly relieved using the so-called phase field method.
It is the purpose of this section to develop a
phase field model for the step problem. We shall demonstrate
that this method provides a powerful tool
permitting  detailed study of time-dependent step kinetics.

The crux of the phase field approach\Ref\pfield{
The idea of phase field approach in the crystal growth
context was put forward by many authors, for example,
G. Fix, in {\sl Free Boundary Problems}, edited by A. Fasano
and M. Primicerio,
Pittman, London, 1983; J.B. Collins and H. Levine,
\journal Phys. Rev. B. &31&6119(85); J.S. Langer, in
{\sl Directions in Condensed Matter Physics}, edited
by G. Grinstein and G. Mazenko, World Scientific,
Singapore, 1986.}
lies on the introduction of an
order parameter $\phi\rt$ indicating the phase at a particular
position. In our model,
local stable minima of the order parameter correspond to
terraces whereas rapid spatial variation of the order
parameter locates the position of steps.
Now we introduce a stochastic phase field model in the following
dimensionless form
 $$ \frac{\pa u\rt}{\pa t}=\nabla^2 u -u+\frac{1}{2}\frac{
\pa \phi}{\pa t} +q\rt  \eqn\phaone $$
$$\tau \frac{
\pa \phi\rt}{\pa t}=\xi^2 \nabla^2\phi +a\sin (\pi\phi)
  -u+\Delta +q_{\phi}\rt  \eqn\phatwo $$
where the last equation can be derived through a pure relaxational
dynamics $\tau \pa \phi/ \pa t= -\delta F/\delta\phi
+q_{\phi}$
from the hypothetical free energy
 $$ F\{\phi,u\}=\int d^2{\bm r}\left[
\frac{\xi^2}{2}(\nabla\phi)^2+\frac{a}{\pi}\cos(\pi\phi)
  +(u-\Delta)\phi \right] . \eqno\eq $$
The Langevin noise $q$ satisfies \qvar\ while
$q_{\phi}$ has zero mean and variance
$$ \VEV{q_{\phi}\rt q_{\phi}(\vecr',t')}=
 \frac{2\Gamma\xi^2}{\tau}\delta({\bm r}
     -{\bm r}')\delta(t-t') .  \eqn\qvar $$
The mangitude of variance is so chosen as to reproduce
the correct interface noise $\eta$ in the original sharp interface
model discussed in the previous sections.

Equations (7.1) and (7.2) are extensions of the classical
phase field model for solidification to the present problem
with thermal fluctuations. A sinusoidal potential term
is introduced to facilitate the description of multiple steps,
by identifying the degenerate minima $\phi \sim (2i+1)\pi$
,where $i$ is an arbitrary integer, with terraces.
The location of the moving steps are defined by
the condition $\phi\rt =2i\pi$.
The phase field model captures, phenomenologically,
the effect of line tension and finite attachment kinetics
by having finite values for $\xi$ and $\tau$, since
$\xi$ can be regarded as the thickness of
steps and $\tau$ reflects the rate of response of the phase field.
Parameter $a$ is the strength
of the potential which can be taken as a constant of
order unity, when considering the limit of small
supersaturations.

Formal methods have been used to establish
the phase field model as a proper regularization of the
original sharp interface problem \dimone\ to \dimthree.
It has been shown\rlap,\Ref\caginalp{
G. Caginalp, \journal Phys. Rev. A. &39&5887(89);
a different formal procedure is carried out by
R. Kupferman, O. Shochet, E. Ben-Jacob and
Z. Schuss, \journal Phys. Rev. B. &46&16045(92).}
again formally using matched asymptotic expansions,
that the phase field model recovers the sharp interface
model by taking appropriate limits of quantities
$\tau \to 0, \xi \to 0$, and $a\ll 1$.
Following the standard procedures, it can be
shown for our present model that the value
of $u$ at a step satisfies
 $$ u_{step} \approx \Delta  - d_0 \kappa -\beta v_n
 +\eta      \eqno\eq $$
with the correspondences
$d_0 \sim \xi \sqrt{a}$, $\beta\sim
\tau\sqrt{a}/\xi$, and that the projected Langevin noise $\eta$
is governed by the correlator \qvartwo.

Equations \phaone\ and \phatwo, although stiff for small
parameters $\xi$ and $\tau$, are suitable for
direct numerical analysis\rlap.\Ref\lowen{
For more sophisticated numerical scheme for one dimensional
phase field models, see  H. Lowen, J. Bechhoefer
and L.S. Tuckerman, \journal Phys. Rev. A. &45&2399(92).}
While simulations on the physical two dimensional geometry
is straightforward, we focus here on the deterministic
one dimensional problem, corresponding to straight steps.
First of all, numerical evidence suggests that \phaone\
and \phatwo\ support steady state solutions with
constant velocity, for infinite equidistant step trains
with arbitrary spacing $\lambda$, and at all
values of supersaturation $0<\Delta<1$.
The most interesting application, however,
concerns the time-dependent motion of finite, in general
non-equidistant step trains.
Such a situation is illustrated in Fig.7,
where a snapshot of the adatom density and order parameter
profiles is displayed for a non-equidistant train of six steps.
The parameters used in our numerical calculation are $\Delta=0.1,
a=1, \tau=2.5\times 10^{-3}, \xi=0.05$, and the average
spacing between steps is $\lambda\approx 2.5$.
Strong overlap of the terrace diffusion fields is evident, even for
average step spacing of $\lda\approx 2.5$.
\FIG\seven{Snapshot of: (a) adatom density and, (b) order
parameter profiles, for a train of six steps. $\Delta=0.1$.}
\FIG\eight{Time evolution of a finite step train,
starting from an initial configuration
of six steps evenly spaced at $\lambda=0.5$. In the
phase field ($\phi$)  plot, time sequences are,
from left to right, $t=0.5, 20, 60, 100$ and $140$. For
clarity, only three instances of the diffusion field ($u$)
at $t=0.5, 60$, and $140$ are shown. $\Delta=0.1$}
To dispel a reasonable misgiving, it is not hard to
realize that an effective short range repulsion exists
between steps spaced on the scale of $\xi$. This conveniently
avoids the possibility of steps behind overtaking others in the
front, which in reality would create overhangs and involve
prohibitive energy penalty.

A train of {\it finite} number of steps can not
propagate with equal spacing indefinitely, because
there is no corresponding steady state solution.
In Fig.8, we consider the temporal evolution of
a train of six steps starting from a configuration
of equal step spacing $\lambda=0.5$ at $\Delta=0.1$.
Step configurations at $t=0.5, 20, 60, 100, 140$ are shown
in sequence.  Both the leading step and the trailing step
move faster than the steps inside the train, because each of
them borders on an infinite terrace with richer adatom
supply than other terraces.
Therefore one expects the step train to pile up at the
rear and spread at the front.
This intuitive conclusion is indeed vividly observed in
Fig.8, where we see the bunching of the last three steps
into a triple step, and the continuous breaking away of
the leading steps from the train. Meanwhile, the
overall extension of the step train increases monotonically in time.
At $t=140$, the total length of the train has increased from
an initial value of $3.0$ to $16.14$, by a factor of five.
It is also interesting to monitor the subsequent evolution of
the triple step at the rear in Fig.8.
First it is clear that a step with multiple
height moves much slower than the isolated steps. Consequently,
the triple step lags farther behind the other steps until
the leading step within the bunch finds a larger terrace
in the front and accelerates to break away from the group.
After a further dissociation of the remaining double step,
the triple bunch eventually disassembles into elementary steps
again. The qualitative conclusions we can now draw, that
finite step trains are susceptible to bunching at the rear and
spreading in the front, and that the average step density in
a bunch decreases with time, are consistent with the
analysis of Mullins and Hirth\refmark{\mullinshirth}
and with the general results of Frank's kinematic wave
theory\rlap.\refmark{\kinematic}

In short we see that, the complex dynamics of a finite step
train result from the combined action of two effects --- of step
grouping from the rear and spreading from the front ---
each initiated from the two ends of the step train.
Both effects can nevertheless be more instructively illustrated
when examined separately, in Fig.9 and Fig.10. For instance,
Fig.9 shows the
temporal development of pairwise bunching undulations
on a semi-infinite step train.  Clearly,
the configuration of the step train at any time consists of
a region with disturbances and an unperturbed region, separated
by a propagating ``shock'' front. Within the region of disturbances,
a pairwise bunching of steps occurs. In addition,
the front appears to propagate uniformly with a constant velocity.
A different behavior is observed for a moving semi-infinite step train
which terminates on the right, as shown in Fig.10. Here no
bunching instability is present and the step train
spreads smoothly.
\FIG\nine{Pairwise bunching of steps at the rear of
   a semi-infinite step train with
   uniform initial spacing $\lambda=0.75$ at $\Delta=0.2$.
   Curves from left to right are at $t=24, 48, 72, 96$,
   and $120$. Adatom density profiles
   are vertically displaced by multiples of $0.2$ to show clearly
   the propagation of bunching instability into the step train.}
\FIG\ten{Uniform spreading at the front of a semi-infinite
   step train with constant initial spacing $\lambda=0.75$
   at $\Delta=0.2$. Step configurations from left to right
   correspond to $t=24, 48, 72, 96$, and $120$. Adatom density
   profiles are vertically displaced by multiples of $0.4$
   for clarity.}
It is then inferrable from the foregoing discussions,
confirmed by our numerical simulations,
that whenever a large step train of lower step density
invades that of a higher density, a shock front froms at the
interface which propagates into the train of higher density
and leaves behind a region of bunched step configurations.
For a high density train running into a low density train,
on the other hand, step spacings at the interface
vary smoothly in space and interpolate between values
characteristic of the two step trains.

These one dimensional examples which we have hitherto
considered are used to illustrate the utility of the phase
field model to the study of microscopic kinetics of monoatomic
steps. The major limitation of the present method
lies in its difficulty to be generalized to the case of
asymmetric step kinetics, due to the discontinuity in
adatom density across the steps. Nevertheless,
many physically important factors,
such as the impurity obstruction to step motion, temporal
changes of growth environment, spatial inhomogeneity in
supersaturation, crystalline anisotropy, and
the effect of island nucleation, can all be conveniently
investigated using this phase field approach.
Detailed studies of these aspects, in two dimensions,
will be presented in the future.

\endpage
%\bigskip %\ACK
\ack
We are very grateful to James Langer for constant
encouragements and for helpful suggestions.
This work was supported in part by the NSF Grant No. PHY89-04035,
and the NSF Science and Technology Center for
Quantized Electronic Structures (Grant No. DMR91-20007).
\endpage
\refout
\endpage
\figout
\end